\documentclass[epj,nopacs]{svjour}
\usepackage[T1]{fontenc}
\usepackage[latin1]{inputenc}
\usepackage{graphicx}
\usepackage{epsfig}
\usepackage{amssymb}
\usepackage{amsmath}
\usepackage{latexsym}
\newcommand{\centre}[2]{\multispan{#1}{\hfill #2\hfill}}
\newcommand{\dotted}{\protect\mbox{${\mathinner{\cdotp\cdotp\cdotp\cdotp\cdotp\cdotp}}$}}
\newcommand{\dashed}{\protect\mbox{- - - -}}

\newcommand{\chain}{\protect\mbox{--- $\cdot$ ---}}

\newcommand{\full}{\protect\mbox{------}}
\newcommand{\EJP}{{\it Eur. J. Phys.} } 

\newcommand{\jpg}{{\it J. Phys. G: Nucl. Part. Phys.} }   

\newcommand{\NIM}{{\it Nucl. Instrum. Methods\/} }
\newcommand{\NP}{{\it Nucl. Phys.} }
\newcommand{\PL}{{\it Phys. Lett.} }
\newcommand{\PR}{{\it Phys. Rev.} }
\newcommand{\PRL}{{\it Phys. Rev. Lett.} }

\newcommand{\etal}{{et al.\/}}
\begin{document}
\title{The $\theta_{13}$ and $\delta_{CP}$ sensitivities of the SPL-Fréjus project revisited}
\author{Jean Eric Campagne, Antoine Cazes}
\institute{ Laboratoire de l'Accélérateur Linéaire -
Université Paris-Sud - B\^at. 200 - BP 34 -
91898 Orsay Cedex, France}
\mail{campagne@lal.in2p3.fr}
\date{\today}
%
%
%
%
\abstract{
An optimization of the CERN SPL beam line has been performed guided by the sensitivities to the $\theta_{13}$ mixing angle and to the $\delta_{CP}$ Dirac CP violating phase. A UNO-like 440 ktons water \v{C}erenkov detector located at 130~km from the target in a new foreseen Fréjus laboratory has been used as a generic detector. Concerning the $\delta_{CP}$ independent $\theta_{13}$ sensitivity, a gain of about $20\%$ may be reached using a $3.5$~GeV proton beam with a 40~m long, 2~m radius decay tunnel compared to the up to now considered $2.2$~GeV beam energy and 20~m long, 1~m radius decay tunnel. This may motivate new machine developments to upgrade the nominal SPL proton beam energy.}

\maketitle

\section{Introduction}
The very near future of the neutrino long baseline experiments is devoted to the study of the oscillation mechanism in the range of $\Delta m^2 = \Delta m^2_{atm} \approx 2.4\times10^{-3}\mathrm{eV}^2$ \cite{SKNU04,K2KNU04}
 using conventional $\nu_\mu$ beams. The current K2K experiment in Japan \cite{K2KNU04}, and the forthcoming MINOS in the USA \cite{MINOS} take benefit of low energy beam to measure the $\Delta m^2$ parameter using the disappearance mode $\nu_\mu\rightarrow\nu_\mu$, while OPERA/ICARUS experiments \cite{OPERA,ICARUS} using the high energy CNGS beam \cite{CNGS} will be able to detect $\nu_\tau$ appearance. If we do not consider the LSND anomaly \cite{LSND} that will be further studied soon by MiniBooNE experiment \cite{MINIBOONE}, the three flavor family scenario will be confirmed and accommodated by a $3\times 3$ Pontecorvo-Maki-Nakagawa-Sakata (PMNS) mixing matrix \cite{PMNS} with three angles ($\theta_{12}$,$\theta_{13}$,$\theta_{23}$) and one Dirac CP phase $\delta_{CP}$.  

Beyond this medium term plan, two of the next future tasks of neutrino physics are to improve the sensitivity of the last unknown mixing angle parameter, the so-called $\theta_{13}$, and to explore the CP violation mechanism in the leptonic sector. The present upper bound on $\theta_{13}$ is $\sin^22\theta_{13}<0.14$ for $\Delta m^2 = \Delta m^2_{atm}$ ($90\%$~CL) \cite{CHOOZ}. This sensitivity can be improved using reactor and accelerator experiments. In reactor experiments, one uses $\bar{\nu}_e$ in disappearance mode and may reach $\sin^22\theta_{13}<0.024$ for $\Delta m^2 = \Delta m^2_{atm}$ ($90\%$CL) \cite{Wpaper}. In accelerator experiments, one can use $\nu_e$ and $\bar{\nu}_e$ from $\beta$ beams \cite{BETABEAM} in both  disappearance and appearance modes (\textit{i.e.} $\stackrel{\scriptscriptstyle (-)}{\nu}_e\rightarrow \stackrel{\scriptscriptstyle(-)}{\nu}_\mu$), and also $\stackrel{\scriptscriptstyle(-)}{\nu}_\mu$ in appearance mode (\textit{i.e.} $\stackrel{\scriptscriptstyle(-)}{\nu}_\mu\rightarrow\stackrel{\scriptscriptstyle(-)}{\nu}_e$) with conventional beams either with sub-mega watt proton drivers \cite{NOVA,T2K} or with multi-mega watt proton drivers \cite{T2K,BNLHS,CERN}. The later neutrino beam type, called Superbeam, is foreseen to be extended to produce $\nu_\mu$ beam and $\bar{\nu}_\mu$ beam from muon decays, the so-called Neutrino Factory, in order to study the eventual leptonic CP violation. Such neutrino complex is under study in Japan, in USA and also in Europe at CERN and details may be found in reference \cite{CERN}. A comparison of the performances of $\beta$ beam and Superbeam may be found for instance in reference \cite{DONINI}.  The reactor experiment result on $\theta_{13}$ is straight forward as compared to Superbeam and Neutrino Factory results that are on one hand richer but in an other hand more complex to analyse due to the interplay between the different physics factors $\theta_{13}$, $\delta_{CP}$, sign$(\Delta m^2_{23})$, sign$(\tan(2\theta_{23}))$ \cite{DONINI,DOUBLE-CHOOZ}. 

\begin{figure}
\centering
\includegraphics[width=85mm]{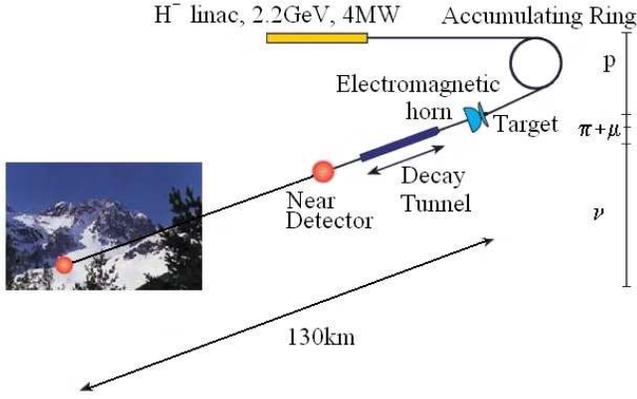}
\caption{\label{fig:Sbeam}Sketch of the SPL neutrino Superbeam from CERN to the Fréjus tunnel.}	
\end{figure}

This paper presents results of a new simulation of the SPL (Super Proton Linac) Superbeam that could take place at CERN \cite{SPL}, using for definitiveness a UNO-like 440kT fiducial water \v{C}erenkov detector \cite{UNO} located in a new enlarged underground laboratory under study in the Fréjus tunnel, $130$~km away from CERN \cite{mosca}. The SPL neutrino beam is created by  decays of pions, muons and kaons produced by the interactions of a $4$~MW proton beam impinging a liquid mercury jet \cite{CERN}. Pions, muons and kaons are collected using two concentric electromagnetic lenses (horns), the inner one and the outer one are hereafter called "Horn" and "Reflector" respectively \cite{Meer}. The horns are followed by a decay tunnel where most of the neutrinos are produced. A sketch of the beam line is shown on figure~\ref{fig:Sbeam}.  

The analysis chain consists of different stages: the simulation of the interactions between the proton beam and the mercury target, the propagation of the resulting secondary particles through the magnetic field and the materials of the horns, the tracking of $\pi^\pm$, $K^{\pm,0}$ and $\mu^\pm$ until they decay, the computation of the neutrino flux at the detector site, and finally the statistical analysis. A part of the simulation chain has already been described in reference \cite{nuFact134,MMWPSCazes}.

Compared to recent papers on the same subject \cite{DONINI,JJG,Mezzetto}, we have reoptimized the Horn and Reflector shapes \cite{nuFact138}, and introduced the kaon background simulation which allows us to update the SPL beam energy. The organization of this document follows the simulation chain: the interaction between the proton beam and the mercury target is presented in the second section. The kaon production is detailed in the third section. The simulation of the horns is described in the fourth section, while the algorithms used to compute the neutrino fluxes are explained in the fifth section. Then, the sensitivities to $\theta_{13}$ and $\delta_{CP}$ are revisited with new studies about the optimization of the proton beam energy, the pion collection and the decay tunnel geometry.

\section{Target simulation}
\label{sec:target}
Since hadronic processes are crucial to describe the interactions of the proton beam on the target, the FLUKA simulator \cite{fluka} has been chosen for this first step of the simulation. 
\begin{table}
\centering
\caption{\label{tab:targ}Liquid mercury jet parameters.}
\begin{tabular}{ll}
\hline\noalign{\smallskip}
    \centre{2}{Hg target}\\
\noalign{\smallskip}\hline\noalign{\smallskip}
      Hg jet speed & $20$~m/s \\
      density & $13.546$ \\
      Length, radius & $30$~cm, $7.5$~mm \\
 \noalign{\smallskip}\hline
\end{tabular}
\end{table}
The target used in the present study is a mercury liquid jet \cite{CERN} simulated by a cylinder $30$~cm long (representing two hadronic lengths) and $1.5$~cm diameter (see table~\ref{tab:targ}). Other types of target are under study \cite{CERN}. The pencil like simulated proton beam is composed of $10^6$ mono-energetic protons. The beam axis is also the symmetry axis of the target and the horns and the decay tunnel. Simulations have been performed for $2.2$~GeV proton kinetic energy, the up to now nominal design \cite{SPL}, as well as for $3.5$~GeV, $4.5$~GeV, $6.5$~GeV and $8$~GeV according to possible new designs \cite{MMWPSGaroby}.

\begin{figure*}
\centering
\includegraphics[height=95mm]{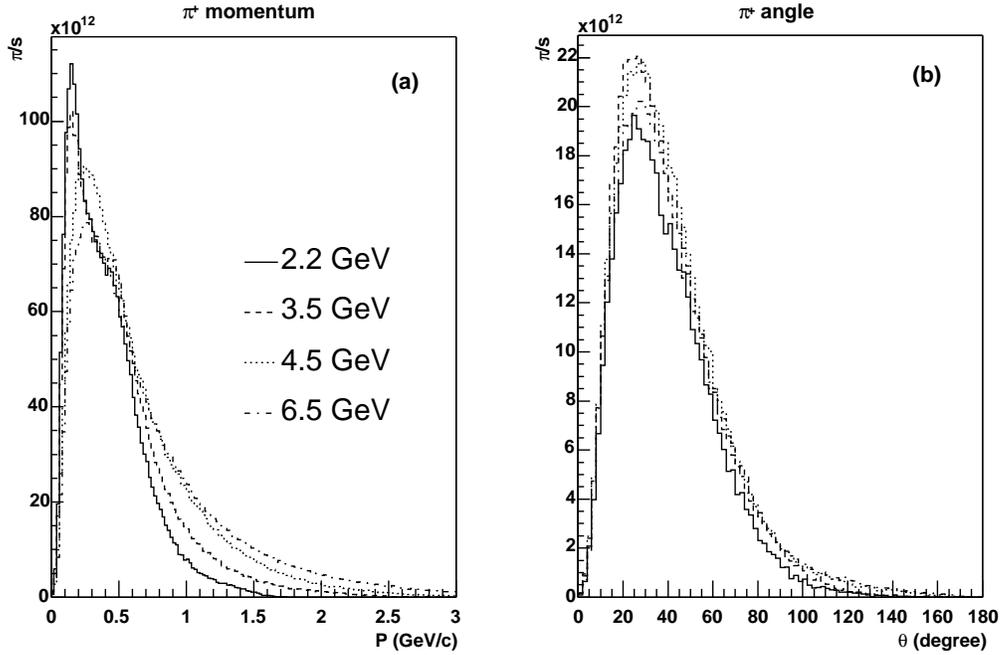}
\caption{\label{fig:compEner2} (a) $\pi^+$ momentum distribution per second at the exit of the target for the different proton beam energies studied, simulated with FLUKA, and (b) $\pi^+$ angle with respect to the beam axis of the pion having a momentum between $0.5$~GeV/c and $0.7$~GeV/c. The different SPL beam kinetic energies presented are (\full) $2.2$~GeV, (\dashed) $3.5$~GeV, (\dotted) $4.5$~GeV and (\chain) $6.5$~GeV.}
\end{figure*}

Particle production yields are summarized in table~\ref{tab:nbPart}. The pion momentum spectra obtained at different energies and normalized to a 4MW SPL beam power are presented in figure~\ref{fig:compEner2}(a). At low energy, pions come from $\Delta$ decays while the high energy part is dominated with multi pion production. At very low energy, for $P<200$~MeV/c, pions come from $\Delta$ produced by protons of the target excited by the beam interactions, while for  higher energy, pion production is due to transformation of protons of the beam into $\Delta$.

The horns are designed to focus the $600$~MeV/c pions (see section~\ref{sec:horn}) and the variation of the number of such pion is rather smooth with respect to the beam energy considering a 4MW fixed beam power: $4.19\times10^{13}\pi$/s for the $2.2$~GeV beam, $4.91\times10^{13}\pi$/s for the $3.5$~GeV beam, $5.14\times10^{13}\pi$/s for the $4.5$~GeV beam, and $4.92\times10^{13}\pi$/s for the $6.5$~GeV beam. The main difference is made by the angular distribution. Figure~\ref{fig:compEner2}(b) shows this distribution for the $\pi^+$ exiting the target with a momentum between $500$~MeV/c and $700$~MeV/c. The acceptance of the horns is limited to the pion below $25^\circ$, and we see that more pions are accepted by the horns for the $3.5$~GeV and $4.5$~GeV proton beams compared to other beam energies.

\begin{figure*}
\centering
\includegraphics[height=95mm]{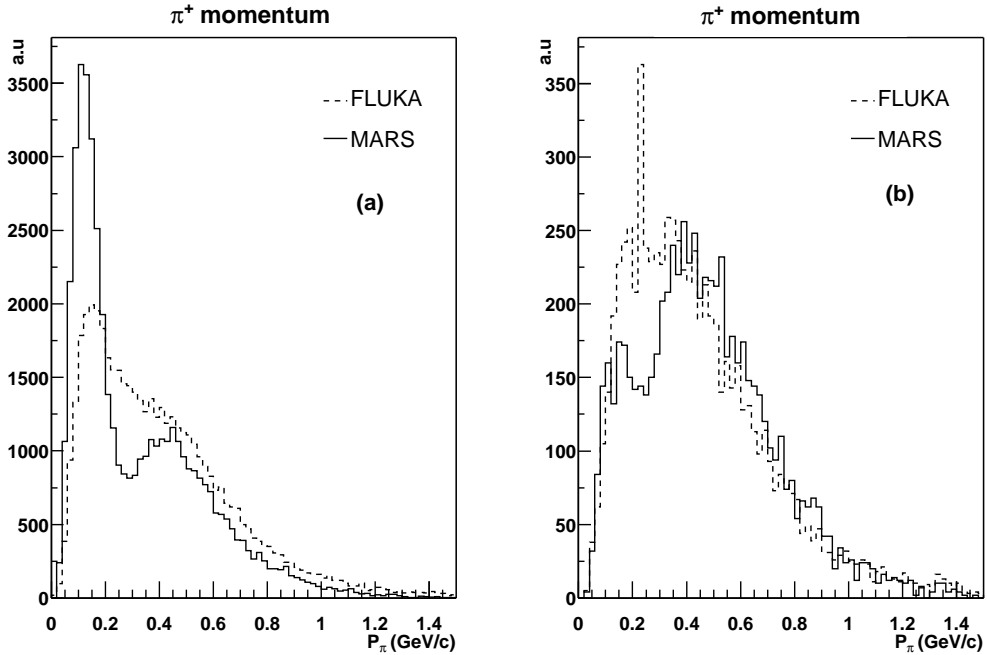}
\caption{\label{fig:compFlukaMars}$\pi^+$ momentum distribution at the exit of the target (a) and at the exit of the horns (b), simulated by FLUKA (\dashed) and by MARS (\full).}
\end{figure*}

The secondary proton and neutron rates induce important radiation damages and power dissipation in the horns which have been addressed in reference \cite{nuFact134}, and which will require specific R\&D effort. At $2.2$~GeV, kaon yields are very low, but it has a dramatic energy dependence as further studied in section~\ref{sec:kaon}. It is worth  mentioning that the numbers in table~\ref{tab:nbPart} are not to be taken as face values, because the cross sections of pion and kaon productions using proton beam are still under studies as for instance by the HARP experiment \cite{harp}.
\begin{table*}
\centering
\caption{\label{tab:nbPart}Average numbers of the most relevant secondary particles exiting the $30$~cm long, $1.5$~cm diameter mercury target per incident proton (FLUKA). The $\mu^+/\mu^-$ numbers and the $K^+/K^0$ numbers have been multiplied by $10^4$. Note that the $K^-$ production rate is at the level of $10^{-5}$ per incident proton.}
\begin{tabular}{@{}l*{15}{l}}
\hline\noalign{\smallskip}
$E_k$ (GeV) & p & n & $\gamma$ & $e^+$ & $e^-$ & $\pi^+$ & $\pi^-$ & $\mu^+$ & $\mu^-$ & $K^+$ & $K^0$ \\
\noalign{\smallskip}\hline\noalign{\smallskip}
$2.2$ & $1.4$     &  $17$   & $5.0$  &   $0.08$  &  $0.17$    &  $0.24$   &  $0.18$ & $4$ & $1$ & $7$ & $6$ \\

$3.5$ & $1.8$    &  $23$  & $7.0$   &  $0.15$  &  $0.28$    &  $0.41$   &  $0.37$ & $10$ & $3$ & $35$ & $30$ \\

$4.5$ &  $2.3$  &  $25$ & $7.7$  & $0.21$  &  $0.35$    &  $0.57$ & $0.39$&  $11$ & $3.3$  &  $93$ &  $68$       \\

$8$   &  $3.1$  &  $33$ & $11.0$ & $0.41$ & $0.63$    &  $1.00$  &  $0.85$ &  $30$ &  $9.5$ &  $413$ &    $340$         \\
 \noalign{\smallskip}\hline
\end{tabular}
\end{table*}
The cross section uncertainties are the main source of discrepancy between simulator programs. Some comparisons between FLUKA and MARS \cite{MARS} have already been presented in the same context \cite{nuFact134}. The energy distribution of the pions exiting the target, computed with the two simulator programs FLUKA and MARS, is shown on figure~\ref{fig:compFlukaMars}(a).  The discrepancy is quite large for the low energy part. However, the horns are designed to focus the high energy part of the spectrum (see section~\ref{sec:horn}), and therefore, MARS and FLUKA are in better agreement for the energy spectrum computed at the entrance of the decay tunnel, as shows figure~\ref{fig:compFlukaMars}(b). So,  the discrepancy  at low energy between MARS and FLUKA does not matter too much for the present application. A difference of $10\%$ has been found between the  $\theta_{13}$ sensitivity computed with the two generators (see section~\ref{sec:results}), that can be taken as systematic error. 
\section{Kaon production}
\label{sec:kaon}
The possibility to increase the SPL energy in order to study the optimization of the physics program has been recently pointed out \cite{MMWPSGaroby}. Then, the kaon production should be clearly addressed because it is a source of $\nu_e$ and $\bar{\nu}_e$ background events. The kaon decay channels and branching ratios are presented in table~\ref{tab:BRKP0SL} in \ref{sec:kaons}.

\begin{figure*}
\centering
\includegraphics[height=95mm]{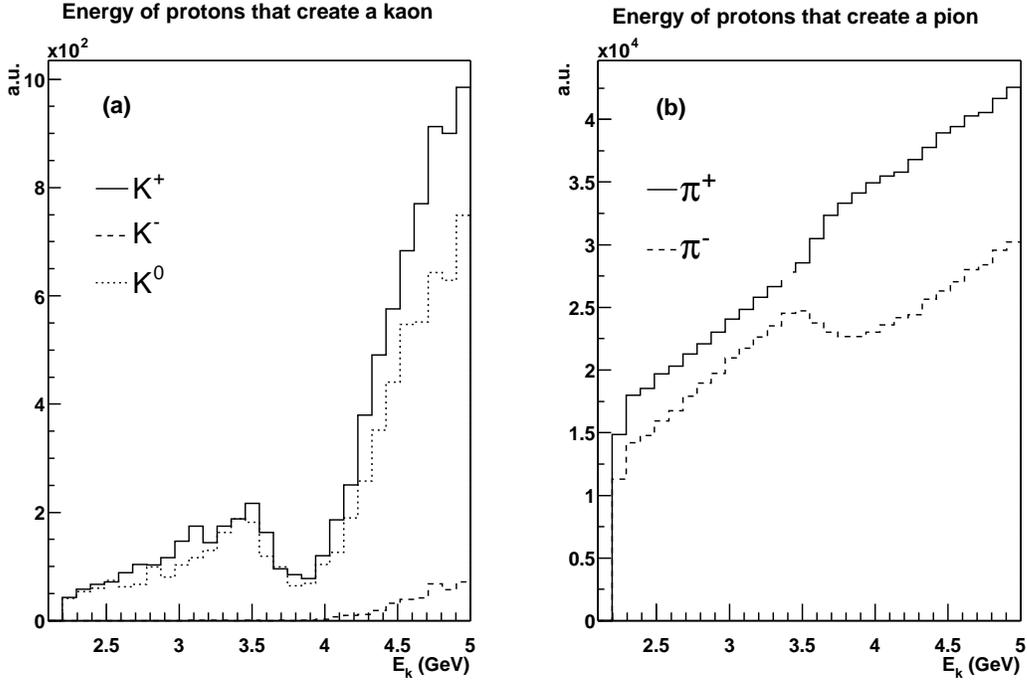}
\caption{\label{fig:KaonsPions}Kaon production (a) as a function of the incident proton beam kinetic energy ($E_k$) for $10^6$ incident protons with (\full) curve for $K^+$, (\dashed) curve for $K^-$ and (\dotted) curve for $K^0$. Pion production (b) in the same conditions with (\full) curve for $\pi^+$ and (\dashed) curve for $\pi^-$.}
\end{figure*}

The target simulation described in section~\ref{sec:target} has been used with $10^6$ p.o.t with kinetic energy uniformly distributed between $2.2$~GeV and $5$~GeV. The momenta of outgoing pions and kaons are recorded when they exit the target. The number of produced $K^{o,\pm}$ at different proton beam energies are presented on figure~\ref{fig:KaonsPions}(a). On the one hand the $K^o$ production rate is similar to the $K^+$ production rate, but  on the other hand the $K^-$ production rate is almost forty times smaller. In comparison, the numbers of $\pi^+$ and $\pi^-$  produced in the same conditions are presented on figure~\ref{fig:KaonsPions}(b). Pion production rate is about two orders of magnitude greater than the kaon production rate.
The behavior of the two pion and kaon production rates are quite different. The $\pi^+$ yield grows smoothly with the proton energy while the production of kaons seems to have two origins, which has been confirmed by FLUKA's authors \cite{FLUKAprivate}. For beam energy below approximatively $4$~GeV, the resonance production model is used, and one notices a low production rate with a maximum at about $3.4$~GeV. For beam energy above  $4$~GeV, the dual parton model is used, and the production rate experiences a threshold effect with a rapid rise. The ratio between positive kaon and pion production rates is about $0.5\%$ between $2.2$~GeV and $4$~GeV and grow up to $2.3\%$ at $5$~GeV. One notices that the transition between the two kaon production models may not be optimal.
\section{Horns simulation}
\label{sec:horn}
The simulation code of the electromagnetic horns is written using GEANT 3.2.1 \cite{geant} for convenience and since electromagnetic processes are dominant, FLUKA has not been considered as mandatory, but this may be revised in a future work. The geometry of the horns has been inspired by an existing CERN prototype and a Reflector design proposed in reference \cite{SIMONE1}. Depending on the current injection, only positive secondary particles or negative secondary particles are focused. The relevant parameters are detailed in table~\ref{tab:specif}.

\begin{figure}
\centering
\includegraphics[width=85mm]{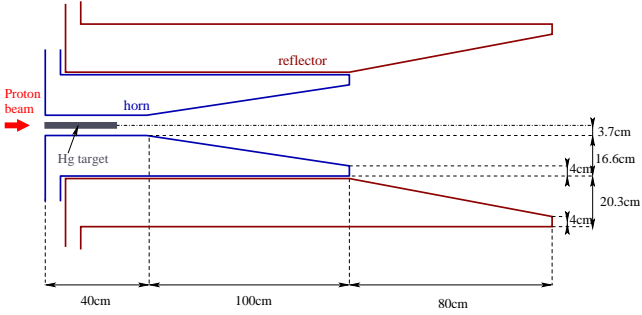}
\caption{\label{fig:plan}Design of the Horn and the Reflector conductors implemented in the GEANT simulation in case of the generation of a 350~MeV neutrino beam. The Hg target is located inside the cylindrical part of the Horn.}
\end{figure}

The mercury target is localized inside the Horn because of the low energy and the large emittance of the secondary pions produced: $$<P_{\pi T}>/<P_\pi> \approx 240~\mathrm{MeV}/400~\mathrm{MeV}$$ ($2.2$~GeV proton beam energy). This explains the Horn design (figure~\ref{fig:plan}), with a cylindrical part around the target, called the neck, which is larger than the transversal size of the target to simulate the room for target handling, and a conic part designed such that the relevant pions are focused as much as possible to exit the magnetic field parallel to the beam axis.

\begin{table}
\centering
\caption{\label{tab:specif}Relevant parameters of the horns in case of the generation of a $260$~MeV neutrino beam (or $350$~MeV in parenthesis). The shapes of the conductors are independent of the proton beam energy, as the focusing has been optimized for a $600$~MeV/c (or $800$~MeV/c) pion momentum.}
\begin{tabular}{@{}l*{15}{l}}
\hline\noalign{\smallskip}
 & Horn & Reflector \\
\noalign{\smallskip}\hline\noalign{\smallskip}
  neck inner radius & $3.7$~cm &  $20.3$~cm \\
  neck length & $40$~cm        &  $120 (140)$~cm \\ 
  end cone inner radius & $16$~cm        &  $35.7$~cm \\
  outer radius & $20.3$~cm      & $40$~cm \\ 
  total length & $120 (140)$~cm       & $190 (220)$~cm \\  
  Alu thickness  & $3$~mm       & $3$~mm \\            
  Peak current & $300$~kA       & $600$~kA \\
  Frequency & $50$~Hz            & $50$~Hz \\
 \noalign{\smallskip}\hline
\end{tabular}
\end{table}
The shape of the horn conductors is a crucial point since it determines the energy spectrum of the neutrino at the detector site. The details of the conductor shape optimization for the present context may be found in reference \cite{nuFact138}. We just recall here some ingredients. For a $\theta_{13}$ driven $\nu_\mu \rightarrow \nu_e$ oscillation, a $\Delta m^2_{23}$ parameter value of $2.5\times10^{-3}\mathrm{eV}^2$, and a baseline distance of $130$~km, the first oscillation probability maximum occurs for a neutrino energy of $260$~MeV. The   optimization of the physics potential depends at first approximation on the pion neutrino characteristics, which energy is fully determined by the pion 2-body decay and boost. To reach an energy of $260$~MeV, the pion needs a $\beta=0.97$, which in turn induces a pion momentum of $600$~MeV/c. Then, the shape of the conic part of the horns is determined such that these $600$~MeV/c pions exit parallel to the beam axis. 

An other shape of the horn conductors has been used to produce a $350$~MeV neutrino beam to compare the sensitivity potential (see section \ref{sec:results}). In that case, keeping the current intensity unchanged ($300/600$~kA), the lengths of the Horn and the Reflector should be increased by $16$\% and $18.5$\%, respectively (see table~\ref{tab:specif}).

Before closing this section, it is worth quoting that the Horn/Reflector conductor shapes optimized in the present study to focus a given pion momentum value, is not affected at first order by a proton beam energy change. What is affected is the production rate of the relevant pions. This Horn/Reflector design consideration would be different if one wished to focus as much as possible all the pions produced for which the mean energy is of course affected by a proton beam energy change.    
\section{Particle decay treatment and flux calculation}
The decay tunnel representation is a simple cylinder with variable length ($L_T$) and radius ($R_T$) filled with "vacuum" and located right after the horns. The default design is a $20$~m long and $1$~m radius cylinder, but simulations have also been conducted with lengths of $10$~m, $40$~m and $60$~m, and radius of $1.5$~m and $2$~m in the spirit of reference \cite{donega}. In the GEANT simulation, to gain in CPU time, only pions, muons and kaons are tracked in the volume of the tunnel, and all particles exiting this volume are discarded. 

Beyond the $1/L^2$ solid angle factor due to the source-detector distance ($L$) which decreases dramatically the fluxes, the neutrino beam focusing is very limited due to the small pion boost factor ($\approx 4$). Therefore, computational algorithms have been used to avoid a too prohibitive CPU time resulting from the simulation of each secondary particle decay. Otherwise, about $10^{15}$ p.o.t would have been necessary to obtain reliable statistics for the estimation of the $\bar{\nu}_e$ flux for instance.

It is worth  pointing out that the particle decays occurring before the entrance of the decay tunnel are also taken into account and treated in the same manner, which is not the case in reference \cite{donega}. 
\subsection{Algorithm description}
\label{sec:algo}
The decay code has been included in the GEANT code. The basic idea of this algorithm is to compute the neutrino fluxes using the probability of reaching the detector for each neutrino produced by a $\pi$ or a $K$ or a $\mu$ particle (on-axis neutrino beam). This method has already been used in reference \cite{donega} and has been modified and extended to the kaon decay chain for the present study. 

Muon neutrino comes mostly from pion decay. In a first stage, each pion is tracked by GEANT until it decays. Then, the probability for the produced muon neutrino to reach the detector is computed. The flux is obtained applying the probability as a weight for each neutrino. All the pions produced in the simulation are therefore useful to compute the flux, and this allows to reduce the number of events in the simulation to $10^6$ p.o.t. In this computation, the decay region (horns and tunnel) is considered as point like compared to the source-detector distance. 

The same method is applied for neutrino coming from muons and kaons with some modifications  because most of the muons do not decay, and there are very few kaons produced (see table~\ref{tab:nbPart}).
The probability computation is presented in appendix~\ref{sec:decayprobcomp}. 
\begin{table}
\centering
\caption{\label{tab:proton}Number of protons on target for different beam energy at 4~MW constant power. One year is defined as $10^7$~s.}
\begin{tabular}{@{}l*{15}{l}}
\hline\noalign{\smallskip}
			Beam energy  & Number of proton \\
			  (GeV)      & per year ($10^{23}$ p.o.t/y) \\
\noalign{\smallskip}\hline\noalign{\smallskip}
			2.2 & 1.10 \\
			3.5 & 0.70 \\
			4.5 & 0.56 \\
			6.5 & 0.40 \\
			8.0 & 0.30 \\
 \noalign{\smallskip}\hline
\end{tabular}
\end{table}
\begin{table*}
\centering
\caption{\label{tab:speciesfluxes}Integral of the total flux of the different species with different settings. The $\nu_\mu$ and $\bar{\nu}_\mu$ fluxes are expressed in $10^{13}/100\mathrm{m}^2/y$ unit while the $\nu_e$ and $\bar{\nu}_e$ fluxes are expressed in $10^{11}/100\mathrm{m}^2/y$ unit. The positive focusing  and negative focusing  are distinguished by a ($+$) sign and a ($-$) sign, respectively. The settings used correspond to different values of $L_T$ and $R_T$, the length and radius of the decay tunnel. Setting (1) means $L_T = 10$~m and $R_T = 1$~m. Setting (2) is the default option and means $L_T = 20$~m and $R_T = 1$~m. Setting (3) means $L_T = 20$~m and $R_T = 1.5$~m. Setting (4) means $L_T = 30$~m and $R_T = 1$~m. Setting (5) means $L_T = 40$~m and $R_T = 1$~m. Setting (6) means $L_T = 40$~m and $R_T = 1.5$~m. Setting (7) means $L_T = 40$~m and $R_T = 2$~m.  Setting (8) means $L_T = 60$~m and $R_T = 1$~m, and finally, setting (9) means $L_T = 60$~m and $R_T = 1.5$~m.}
\begin{tabular}{@{}l*{15}{l}}
\hline\noalign{\smallskip}

			Settings  & \centre{2}{$\nu_\mu$} & \centre{2}{$\nu_e$} 
								&	\centre{2}{$\bar{\nu}_\mu$} & \centre{2}{$\bar{\nu}_e$} \\
								\cline{2-9}
								& $+$ & $-$ & $+$ & $-$ & $+$ & $-$ 
								& $+$ & $-$ \\ 
\noalign{\smallskip}\hline\noalign{\smallskip}

(1): $2.2$~GeV  
	& $5.5$          & $0.4$ 
	& $1.7$          & $0.2$
	& $0.3$          & $4.3$  
	& $0.1$ 				 & $0.8$ \\		

(1): $3.5$~GeV  
	& $7.7$          & $0.7$ 
	& $2.6$          & $0.6$
	& $0.6$          & $6.6$  
	& $0.3$ 				 & $1.3$ \\		
	
(1): $4.5$~GeV  
	& $7.1$          & $1.0$ 
	& $2.8$          & $0.9$
	& $0.5$          & $5.2$  
	& $0.3$ 			   & $1.1$ \\		  

(1): $6.5$~GeV  
	& $8.3$          & $1.2$ 
	& $4.7$          & $1.9$
	& $0.8$          & $5.6$  
	& $0.9$ 			   & $1.8$ \\

(1): $8.0$~GeV  
	& $7.7$          & $1.2$ 
	& $5.1$          & $2.2$
	& $0.9$          & $5.6$  
	& $1.1$ 				 & $2.1$ \\

(2): $2.2$~GeV  
	& $7.6$         & $0.4$ 
	& $3.2$ 				& $0.2$
	& $0.3$ 				& $5.8$  
	& $0.1$ 				& $1.6$ \\		  

(2): $3.5$~GeV  
	& $10.0$          & $0.9$ 
	& $4.4$           & $0.6$
	& $0.7$ 					& $8.5$  
	& $0.3$ 					& $2.2$ \\		  

(2): $4.5$~GeV  
	& $10.9$          & $1.1$ 
	& $5.1$           & $1.0$
	& $0.7$ 					& $6.7$  
	& $0.4$ 					& $1.8$ \\		  

(2): $6.5$~GeV  
	& $10.4$          & $1.4$ 
	& $6.4$           & $2.0$
	& $1.0$           & $7.1$  
	& $0.9$ 					& $2.5$ \\	

(2): $8.0$~GeV  
	& $9.7$          & $1.5$ 
	& $6.7$           & $2.3$
	& $1.2$           & $7.1$  
	& $1.1$ 					& $2.8$ \\

(3): $2.2$~GeV  
	& $9.0$           & $0.6$ 
	& $4.4$           & $0.4$
	& $0.4$           & $6.7$  
	& $0.2$  					& $2.2$ \\

(3): $4.5$~GeV  
	& $13.2$          & $1.5$ 
	& $6.9$           & $1.4$
	& $0.9$           & $8.1$  
	& $0.6$  					& $2.7$ \\

(4): $3.5$~GeV  
	& $10.9$          & $0.9$ 
	& $5.7$           & $0.7$
	& $0.7$           & $9.4$  
	& $0.3$  					& $2.9$ \\

(4): $4.5$~GeV  
	& $11.6$          & $1.2$ 
	& $6.3$           & $1.0$
	& $0.7$           & $7.1$  
	& $0.4$  					& $2.3$ \\

(5): $2.2$~GeV  
	& $8.9$           & $0.5$ 
	& $5.1$           & $0.3$
	& $0.5$           & $6.7$  
	& $0.1$ 					& $2.4$ \\	

(5): $3.5$~GeV  
	& $11.3$          & $0.9$ 
	& $6.5$           & $0.6$
	& $0.8$           & $9.7$  
	& $0.3$ 					& $3.3$ \\	

(5): $4.5$~GeV  
	& $12.3$          & $1.2$ 
	& $7.2$           & $1.0$
	& $0.8$           & $7.5$  
	& $0.4$ 					& $2.6$ \\	

(5): $6.5$~GeV  
	& $11.7$          & $1.6$ 
	& $8.3$           & $2.2$
	& $1.1$           & $8.0$  
	& $0.9$ 					& $3.3$ \\	

(5): $8.0$~GeV  
	& $10.9$          & $1.7$ 
	& $8.5$           & $2.4$
	& $1.3$           & $8.0$  
	& $1.2$ 					& $3.6$ \\	

(6): $3.5$~GeV  
	& $14.5$          & $1.3$ 
	& $10.0$          & $1.0$
	& $1.0$           & $12.3$  
	& $0.5$  					& $5.3$ \\
	
(6): $4.5$~GeV  
	& $15.5$          & $1.7$ 
	& $10.8$          & $1.5$
	& $1.0$           & $9.5$  
	& $0.6$  					& $4.2$ \\
		
(7): $3.5$~GeV  
	& $16.6$          & $1.5$ 
	& $12.9$          & $1.3$
	& $1.3$           & $13.9$  
	& $0.7$  					& $6.9$ \\
	
(7): $4.5$~GeV  
	& $18.2$          & $2.1$ 
	& $14.3$          & $1.9$
	& $1.3$           & $11.1$  
	& $0.8$  					& $5.6$ \\
	
(8): $3.5$~GeV  
	& $11.7$          & $0.9$ 
	& $7.6$           & $0.7$
	& $0.7$          & $10.1$  
	& $0.3$  					& $3.7$ \\
	
(8): $4.5$~GeV  
	& $12.5$          & $1.3$ 
	& $8.1$           & $1.1$
	& $0.7$          & $7.7$  
	& $0.4$  					& $2.9$ \\
	
(9): $3.5$~GeV  
	& $15.1$          & $1.3$ 
	& $12.2$          & $1.0$
	& $1.0$           & $12.8$  
	& $0.5$  					& $6.3$ \\
	
(9): $4.5$~GeV  
	& $16.2$          & $1.8$ 
	& $13.1$          & $1.6$
	& $1.0$           & $9.9$  
	& $0.6$  					& $4.9$ \\
 \noalign{\smallskip}\hline
\end{tabular}
\end{table*}
\subsection{Validation of the algorithm}
The validity of the method presented in the previous section has been tested against a straight forward  algorithm consisting in decaying each pion $N$ times ($N \approx 10^6$) in a full GEANT simulation of the event (decays included). Such method presents the advantage to keep all the information of the neutrino available for further studies. It can be a good approach to compute the muon neutrino flux coming from pion decays. It can also provide the beam profile, but it shows its limits for the muon induced fluxes, especially the $\bar{\nu}_e$ flux. Indeed, this means that each muon is duplicated $N$ times and when a muon decays, it must decay $N$ times again. For $N\approx 10^6$, this is a prohibitive CPU time consuming.
\begin{figure*}
\centering
\includegraphics[height=95mm]{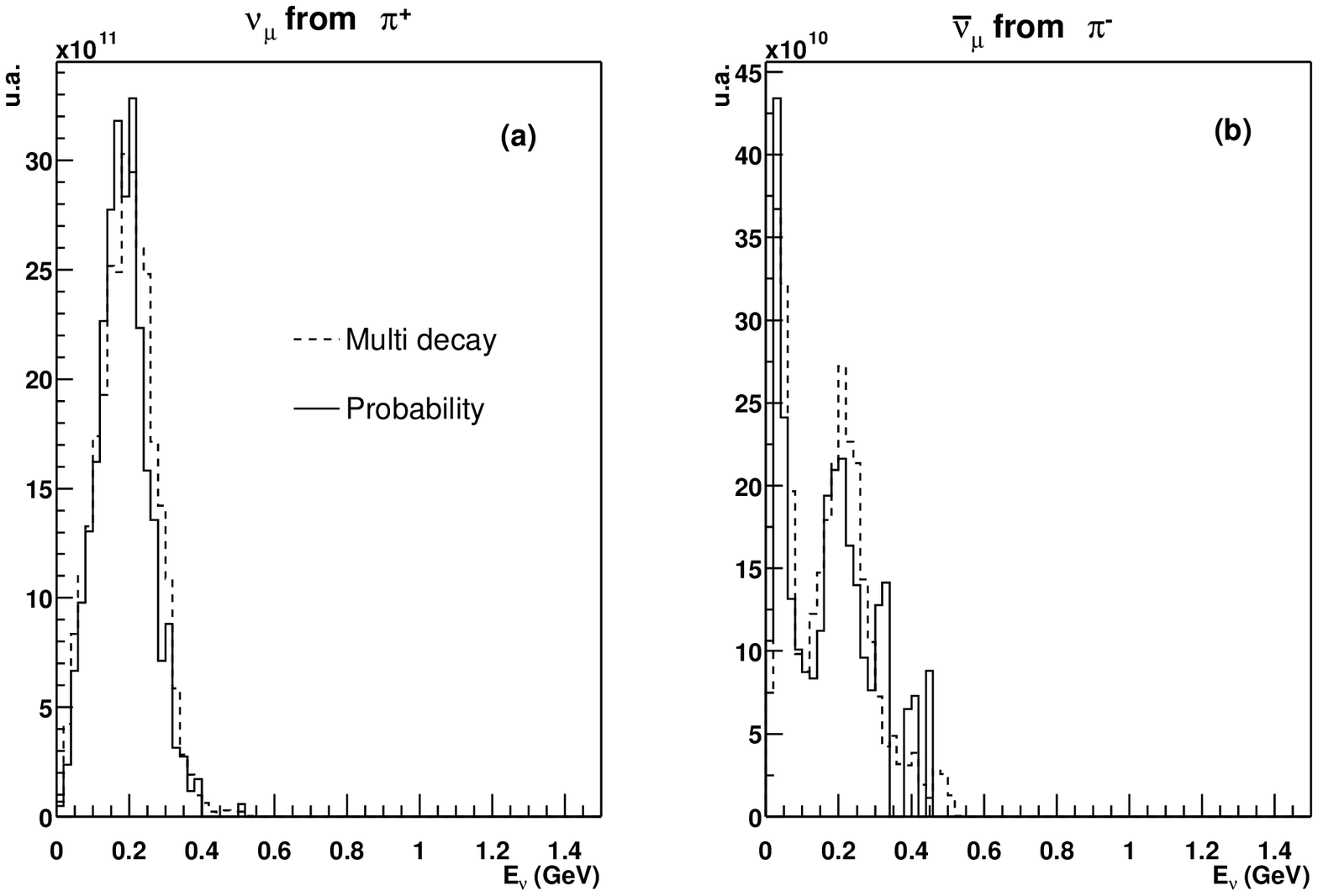}
\caption{\label{fig:compGeantDonega}Comparison between the probability method, (\full) curve, and the full GEANT simulation method, (\dashed) curve, for the $\nu_\mu$ from $\pi^+$ flux (a) and the $\bar{\nu}_\mu$ from $\pi^-$ flux (b). The horns are set to focus positive particles. It should be stressed that the full GEANT simulation has taken roughly 13 times more CPU time than the probability method with the same number of protons on target, and the later simulation is able to produce as well the $\nu_e$ and $\bar{\nu}_e$ fluxes contrary to the former simulation.}
\end{figure*}
The $\nu_\mu$ and $\bar{\nu}_\mu$ fluxes are displayed on figure~\ref{fig:compGeantDonega} for both methods. The two spectra show a clear agreement, and this makes reliable the probability method.
\subsection{Simulated fluxes}
The fluxes are computed at a distance of $100$~km from the source by convention and can be rescaled at any desired distance. They provide the number of the four neutrino species ($\nu_\mu$, $\bar{\nu}_\mu$, $\nu_e$, $\bar{\nu}_e$) passing through a $100$~m$^2$ fiducial area during $1$~year.

In practice, the fluxes are given as a function of the neutrino energy via histograms composed of $20$~MeV bin width.  These histograms are filled with the energy of each neutrino weighted by the probability to reach the detector (section~\ref{sec:algo}). To obtain the fluxes, the histograms are rescaled to the number of p.o.t per year depending on the beam energy. Table~\ref{tab:proton} reports on the number of p.o.t per year for the different energies studied using the definition of one year being $10^7$~s and keeping the beam power constant (\textit{i.e.} $4$~MW).

Three origins are identified in the composition of each neutrino flux:
\begin{description}
	\item[-] neutrinos from pions, which includes neutrinos created by primary pion decays and neutrinos coming from the muons produced by pion decays or muons directly exiting the target. This is the component studied in reference \cite{donega} but with different settings and event generator;
	\item[-] neutrinos emitted during the decay chain of the charged kaons, either by direct production, or produced by the daughter pions and muons;
	\item[-] neutrinos coming from the decay chain of the neutral kaons.
\end{description}

\begin{figure*}
\centering
\includegraphics[height=95mm]{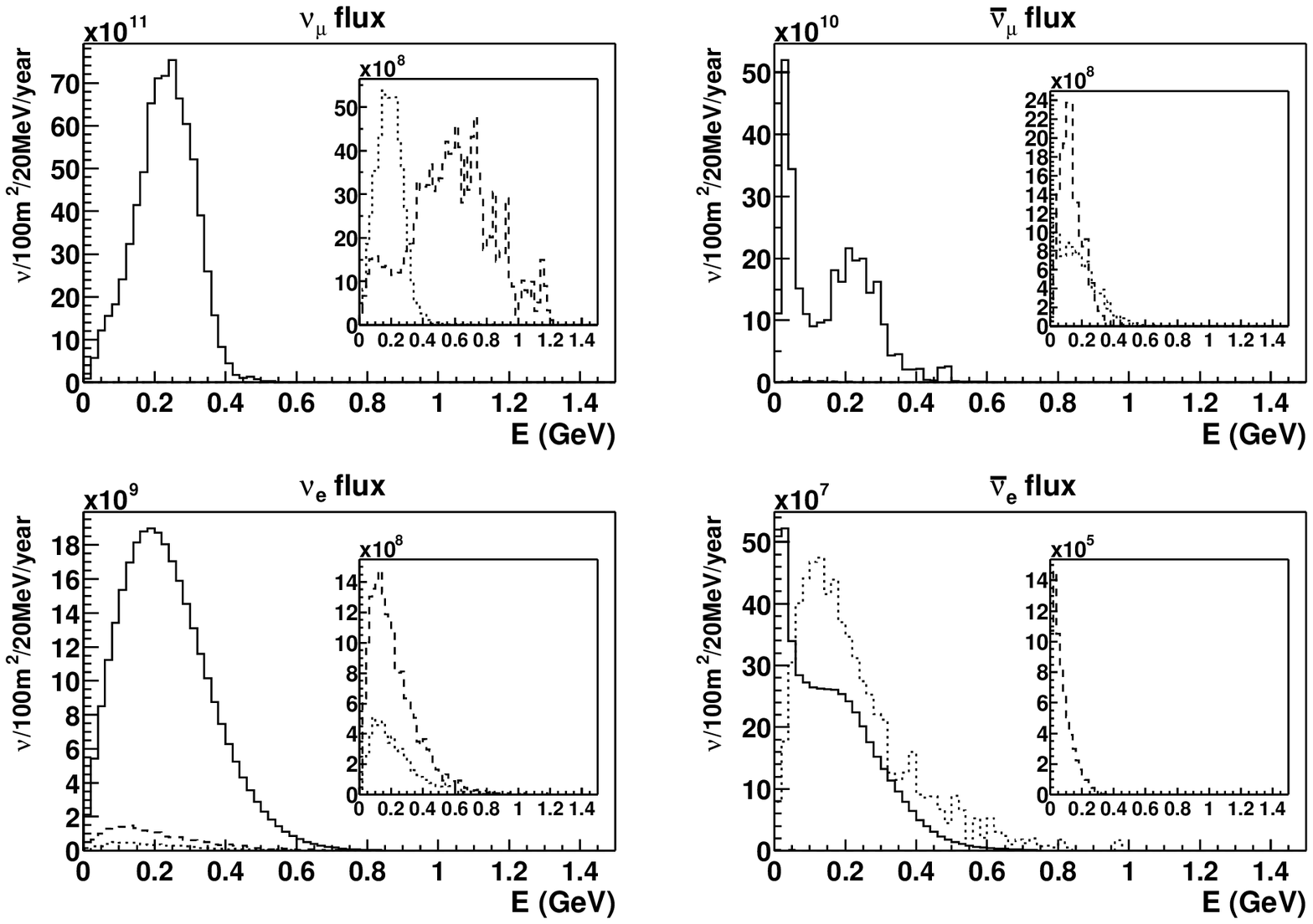}
\caption{	\label{fig:flux22p}Neutrino fluxes, $100$~km from the target and with the horns focusing the positive particles. The fluxes are computed for a SPL proton beam of $2.2$~GeV (4~MW), a decay tunnel with a length of $20$~m and a radius of $1$~m. The top left panel contains the $\nu_\mu$ fluxes, and the top right panel shows the $\bar{\nu}_\mu$ fluxes. The bottom left panel presents the $\nu_e$ fluxes while the bottom right panel displays the $\bar{\nu}_e$ fluxes. The (\full) curve is the contribution from primary pions and the daughter muons, and from primary muons. The (\dashed) curve is the contribution from the charged kaon decay chain, and the (\dotted) curve is the contribution from the $K^0$ decay chain. An insert has been added to the plots to hight light when needed the contribution of charged and neutral kaons.}
\end{figure*}

The three components of the fluxes for the four neutrino species are presented on figure~\ref{fig:flux22p} for positive particle focusing and a proton beam kinetic energy of $2.2$~GeV. The $\nu_\mu$ flux is dominated by the neutrinos of pion decays, but a tail above $500$~MeV (insert on the top left part) is created by the $K^+\rightarrow \mu^+\nu_\mu$ channel, which is anyway at least three order of magnitude below the flux maximum. The $\bar{\nu}_\mu$ flux is mostly due to the decays of $\pi^-$ that are not unfocused by the horns, but the higher energy part comes from $\mu^+$ decays. It is noticeable that the $\nu_e$ and $\bar{\nu}_e$ fluxes are respectively more than $200$ and more than $7000$ times smaller than the $\nu_\mu$ flux. The $\bar{\nu}_e$ are produced in a large part by the $K^0_L\rightarrow\pi^+e^-\bar{\nu}_e$ decay channel and by $\mu^-$ decays, while the  $\nu_e$ flux is dominated by the $\mu^+$ decays.

\begin{figure*}
\centering
\includegraphics[height=95mm]{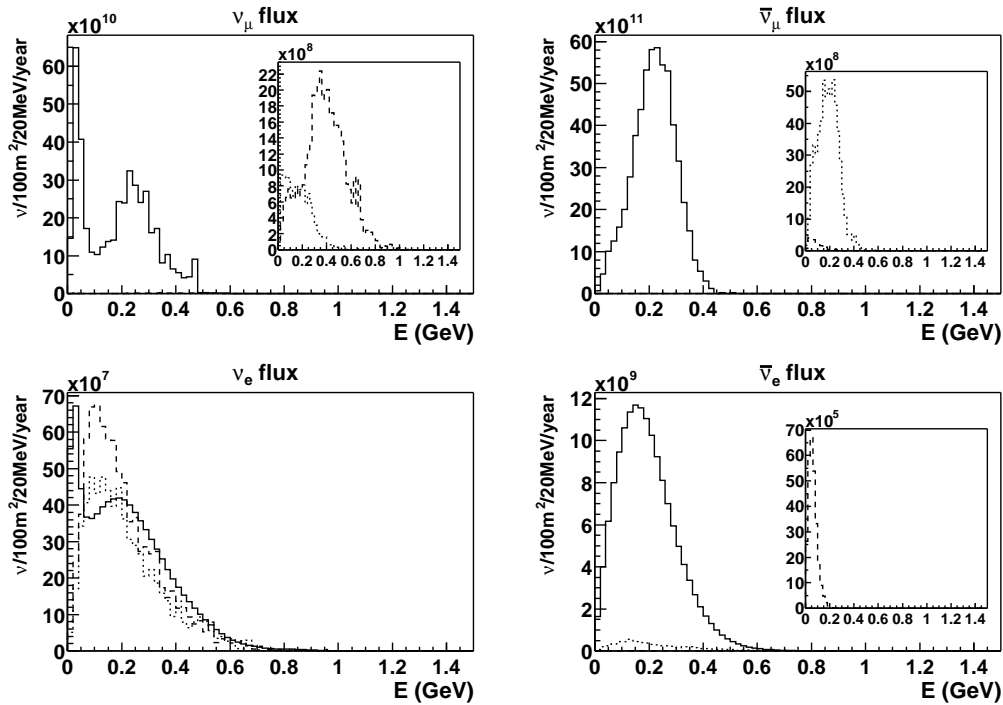}
\caption{\label{fig:flux22m}Same legend as for figure~\ref{fig:flux22p} but the horns are focusing negative particles.}
\end{figure*}

On figure~\ref{fig:flux22m}, the horns are set to focus negative particles keeping other parameters identical. Comparing with positive focusing, one can at first approximation translate the results by exchanging particles and anti-particles, except that the $K^+/K^-$ ratio is about 50 in the beam-target interactions (see table~\ref{tab:nbPart}).

\begin{figure*}
\centering
\includegraphics[height=95mm]{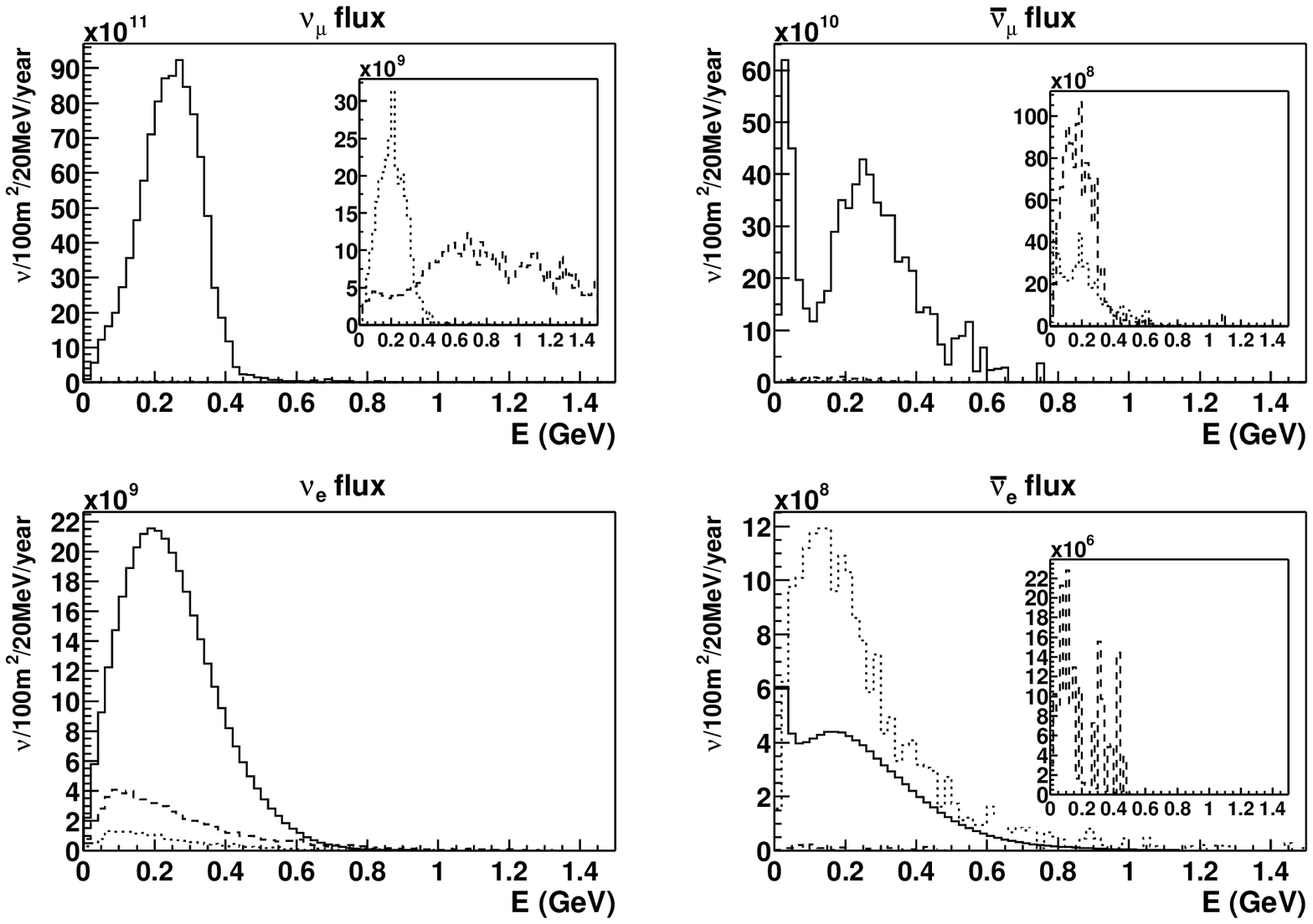}
\caption{\label{fig:flux45p}Same legend as for figure~\ref{fig:flux22p} but for proton beam kinetic energy of $3.5$~GeV (4~MW).}
\end{figure*}

\begin{figure*}
\centering
\includegraphics[height=95mm]{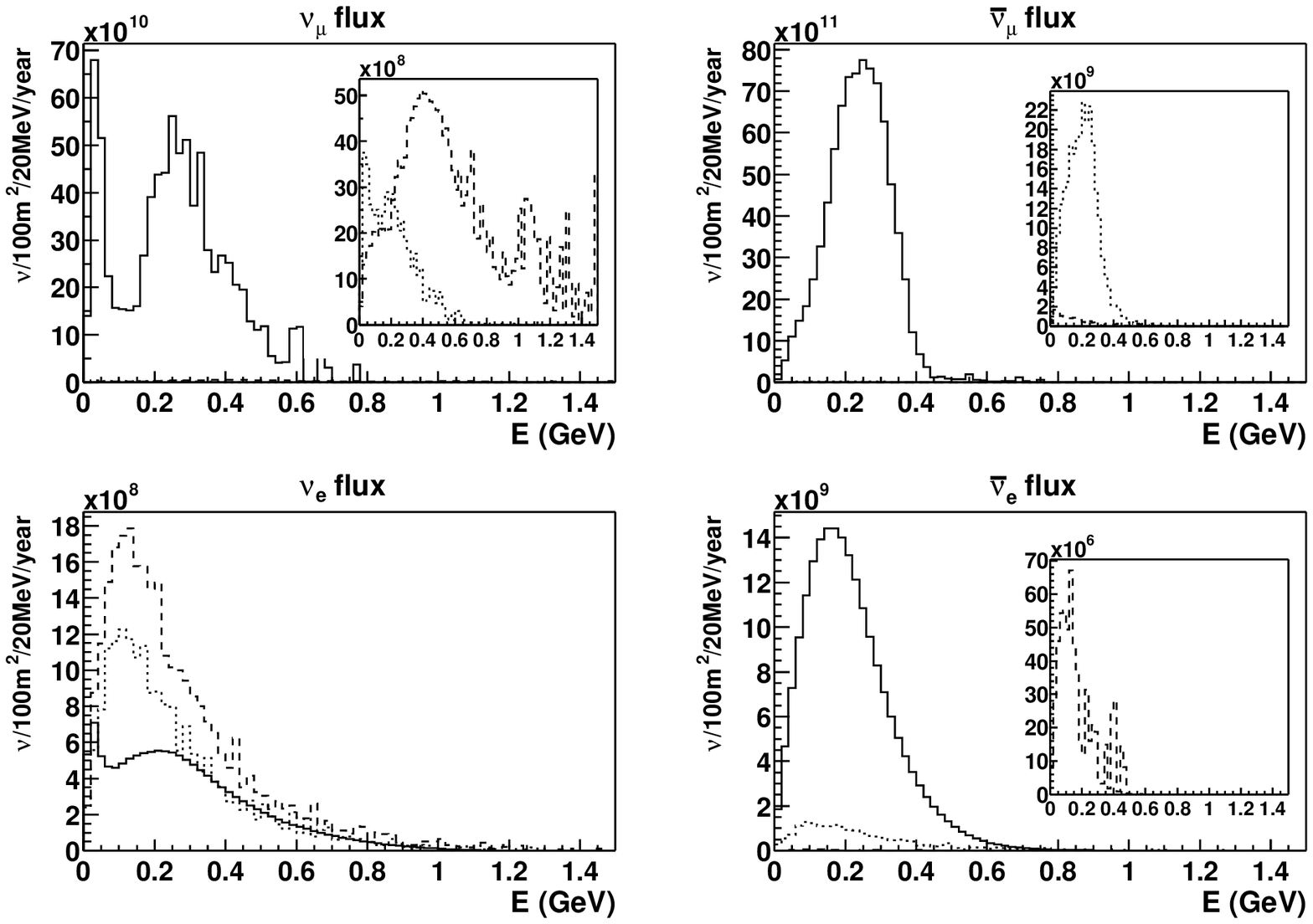}
\caption{\label{fig:flux45m}Same legend as for figure~\ref{fig:flux22m} but for proton beam kinetic energy of $3.5$~GeV (4~MW).}
\end{figure*}

\begin{figure*}
\centering
\includegraphics[height=95mm]{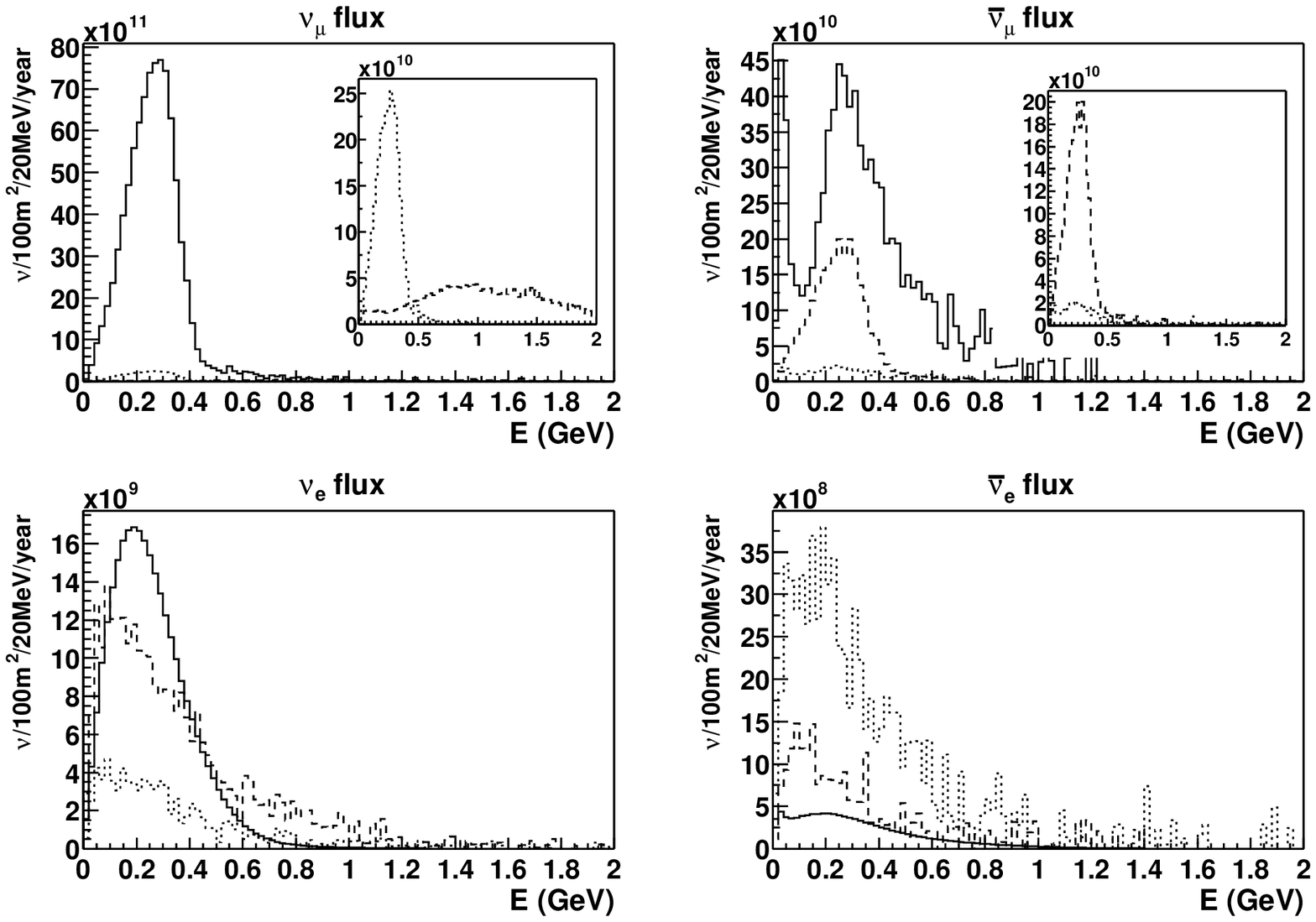}
\caption{\label{fig:flux8p}Same legend as for figure~\ref{fig:flux22p} but for proton beam kinetic energy of $8$~GeV (4~MW).}
\end{figure*}

\begin{figure*}
\centering
\includegraphics[height=95mm]{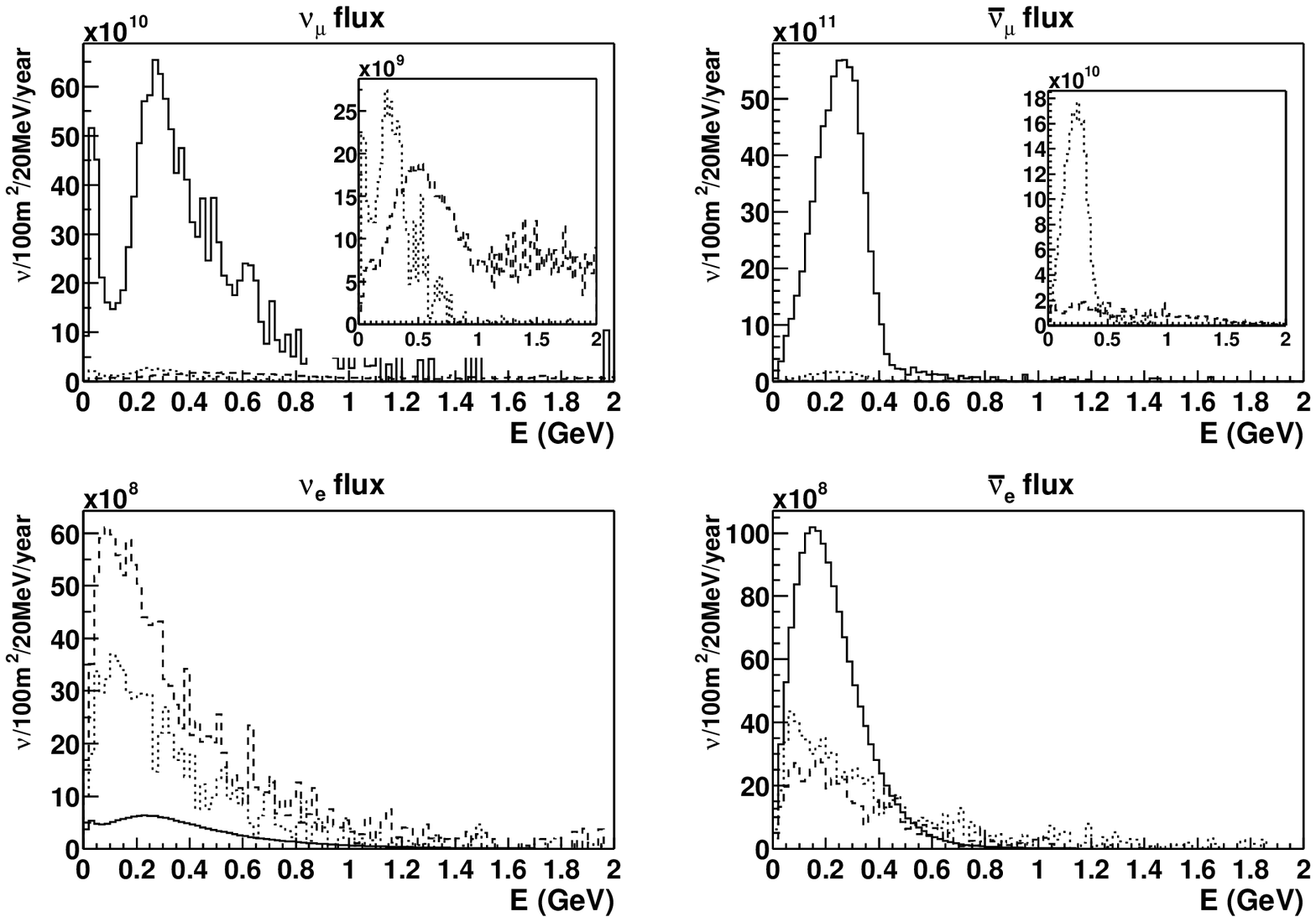}
\caption{\label{fig:flux8m}Same legend as for figure~\ref{fig:flux22m} but for proton beam kinetic energy of $8$~GeV (4~MW).}
\end{figure*}

On figures~\ref{fig:flux45p} and \ref{fig:flux8p}, one observes the evolution of figure~\ref{fig:flux22p} when the proton beam kinetic energy increases to $3.5$~GeV and $8$~GeV, respectively. Correspondingly, the results for negative particle focusing are presented on figures~\ref{fig:flux45m} and \ref{fig:flux8m}. One clearly notices the increase of the kaon induced neutrino contents as the beam energy grows.

On table~\ref{tab:speciesfluxes} are reported the integral of the fluxes when one modifies the decay tunnel length and radius, as well as the beam kinetic energy. Changing the length from $10$~m to $40$~m will increase the $\nu_\mu$ flux by $50\%$ to $70\%$ and in the same time, the number of $\nu_e$ will be multiplied by a factor $1.5$ to $2$. One can notice that going from $40$~m to $60$~m does not increase the signal-like events but increases the background-like events. For a $40$~m length of the decay tunnel, the increase of the radius 
improves the number of signal-like events by $50\%$, and the backgroud increase by $70\%$ to $100\%$. Notice that the $\nu_\mu/\bar{\nu}_\mu$ flux ratio is rather insensitive to the decay tunnel length. The feeling that $L_T = 40$~m and $R_T = 2$~m is a good signal over background compromise is confirmed by sensitivity quantitative studies reported in section~\ref{sec:results}.

Looking at the evolution of $\nu_\mu$ flux with respect to the beam energy, one notices that a maximum is reached around $4.5$~GeV. This is due to the competition between the cross section rise with respect to the energy and the decrease of the number of p.o.t due to the constant SPL power ($4$~MW).
\vspace*{-6mm}
\section{Sensitivity computation ingredients}
The sensitivity to $\theta_{13}$ and $\delta_{CP}$ is computed for a $\nu_\mu\rightarrow\nu_e$ appearance experiment. An analysis program described in reference \cite{MEZZETTONUFACT060} has been used for such sensitivity computation. See table~\ref{tab:param} for the default user parameter values used in this paper. We just remind here some key points of the program. 

It includes a full 3-flavors oscillation probability computation with matter effects, but no ambiguities are taken into account. This latest point may be revisited in a future work using reference \cite{DONINI-2}. Concerning the background events, the $\nu_e/\bar{\nu}_e$ from the beam, the $\nu_\mu e^-$ elastic scattering process, the $\pi^o$ production as well as the $\mu/e$ misidentification are taken into account. The cross-sections from the NUANCE program are used \cite{NUANCE}. The systematics error on the total $\nu_e$ and $\bar{\nu}_e$ fluxes determination is a user parameter and we have used the $2\%$ value considered as a final goal, but also $5\%$ and $10\%$ \cite{MEZZETTONUFACT060}. The detector considered for definitiveness is similar to the UNO detector, \textit{i.e.} a $440$~kt fiducial water \v{C}erenkov detector \cite{UNO}. It is located at $L = 130$~km from CERN, in the foreseen new Fréjus laboratory \cite{mosca}. It is worth  mentioning that if one wants to evaluate the influence of $L$ on the sensitivity, it would mean a re-optimization of the horns for each $L$ envisaged (see section~\ref{sec:horn}).  
\begin{table}
\centering
\caption{\label{tab:param}Default parameters used to compute the sensitivity curves \cite{MEZZETTONUFACT060}. The quoted errors in parenthesis for the $(12)$ and the $(23)$ parameters (absolute value for the masse square differences and relative value for the angles) are coming respectively from the up to date combined Solar and KamLAND results \cite{KAMLAND} and from a 200 ktons-years SPL desappearance exposure \cite{JJG}.}
\begin{tabular}{@{}l*{15}{l}}
\hline\noalign{\smallskip}
			$\Delta m^2_{12} = 8.2 (0.5)\times 10^{-5}~\mathrm{eV}^2$ & $\sin^22\theta_{12} = 0.82 (9\%)$ \\
			$\Delta m^2_{23} = 2.5 (0.1)\times 10^{-3}~\mathrm{eV}^2$ & $\sin^22\theta_{23} = 1.0 (1\%)$ \\
\noalign{\smallskip}\hline\noalign{\smallskip}
		  $L_T = 20$~m & $R_T = 1$~m \\ 
		  $M=440$~kT   & $\epsilon_{syst}=2\%$ \\
\noalign{\smallskip}\hline\noalign{\smallskip}
\centre{2}{Horn/Reflector shapes to produce a $260$~MeV neutrino beam} \\
\noalign{\smallskip}\hline
\end{tabular}
\end{table}
The running time scenario has been fixed either by focusing positive particles during 5 years, either  by focusing positive particles during 1 (or 2) year(s) followed by focusing negative particles during 4 (or 8) years.
\vspace*{-6mm}
\section{Results}
\label{sec:results}
\subsection{The positive only focusing scenario}
The $\theta_{13}$ and $\delta_{CP}$ sensitivities are computed with $\theta_{13} = 0^\circ$ and $\delta_{CP} = 0^\circ$ if not explicitly mentioned. It is worth stressing that the default parameters of table~\ref{tab:param} are used if not contrary mentioned, in particular, the decay tunnel geometry parameters ($L_T=20$~m and $R_T=1$~m), and the horn design to generate a $260$~MeV neutrino beam. 

Table~\ref{tab:nbvsE} presents the number of signal and background events for a $5$ years positive focusing experiment, but with different beam energy settings. The significance parameter is defined in reference \cite{MEZZETTONUFACT060} as\footnote{Contrary to the definition of the significance of reference \cite{MEZZETTONUFACT060}, the systematical factor is applied to the total $\nu_e$ flux in agreement with the sensitivity contour computation.}:
\begin{equation}
\begin{aligned}
\mathcal{S} &= \frac{N^{osc}_{\nu_e}}{\sqrt{ N^{tot}_{\nu_e} + \left(N^{tot}_{\nu_e}\times\epsilon_{syst}\right)^2}}\\
\mathrm{with}\quad N^{tot}_{\nu_e} &= N^{osc}_{\nu_e} + N^{beam}_{\nu_e} + N^{oth. bkg}
\label{eq:significance}
\end{aligned}
\end{equation}
and $N^{osc}_{\nu_e}$ the number of $\nu_e$ events due to $\nu_\mu$ oscillations, $N^{beam}_{\nu_e}$ the number of background events coming from the $\nu_e+\bar{\nu}_e$ contamination of the beam, $N^{oth. bkg}$ the other kinds of background events and $\epsilon_{syst}$ the systematical factor.
%

\begin{table*}
\centering
\caption{\label{tab:nbvsE}Number of events for 5 years positive focusing scenario with default parameters of table~\ref{tab:param}. Other backgrounds are $\pi^0$, $\nu_\mu$-elast., $\mu/e$-missId. The significance parameter is defined by equation~\ref{eq:significance}.}
\begin{tabular}{@{}l*{15}{l}}
\hline\noalign{\smallskip}
		 & $2.2$~GeV & $3.5$~GeV & $4.5$~GeV & $6.5$~GeV & $8$~GeV \\
\noalign{\smallskip}\hline\noalign{\smallskip}
			non oscillated $\nu_\mu$       & $36917$  & $60969$  & $73202$  & $78024$ & $76068$ \\ 
			oscillated $\nu_e$             & $43$     & $60$     & $64$     & $61$    & $56$ \\
			beam $\nu_e$                   & $165$    & $222$    & $242$    & $288$   & $299$ \\
			other background               & $70$     & $105$    & $127$    & $148$   & $152$ \\
			Significance                   & $1.88$   & $2.16$   & $2.17$   & $1.87$  & $1.69$ \\
\noalign{\smallskip}\hline
\end{tabular}
\end{table*}
\begin{table*}
\centering
\caption{\label{tab:thvsE}Minimum $\sin^22\theta_{13}\times 10^3$ in the $(\sin^22\theta_{13},\Delta m^2_{23})$ plane observable at $90\%$ CL computed for different decay tunnel length ($L_T$) and kinetic beam energy ($E_k(proton)$) and 5 year of positive focusing. Other parameters are fixed to default values (table~\ref{tab:param}).}
\begin{tabular}{@{}l*{15}{l}}
\hline\noalign{\smallskip}
		        & $2.2$~GeV & $3.5$~GeV & $4.5$~GeV & $6.5$~GeV & $8$~GeV  \\
\noalign{\smallskip}\hline\noalign{\smallskip}
			$10$~m & $1.10$ & $0.92$ & $1.04$ & $1.07$ & $1.16$ \\
			$20$~m & $1.16$ & $0.92$ & $0.89$ & $1.01$ & $1.12$ \\
			$40$~m & $1.23$ & $1.00$ & $0.99$ & $1.08$ & $1.19$ \\
\noalign{\smallskip}\hline
\end{tabular}
\end{table*}

\begin{figure}
\centering
\includegraphics[width=85mm]{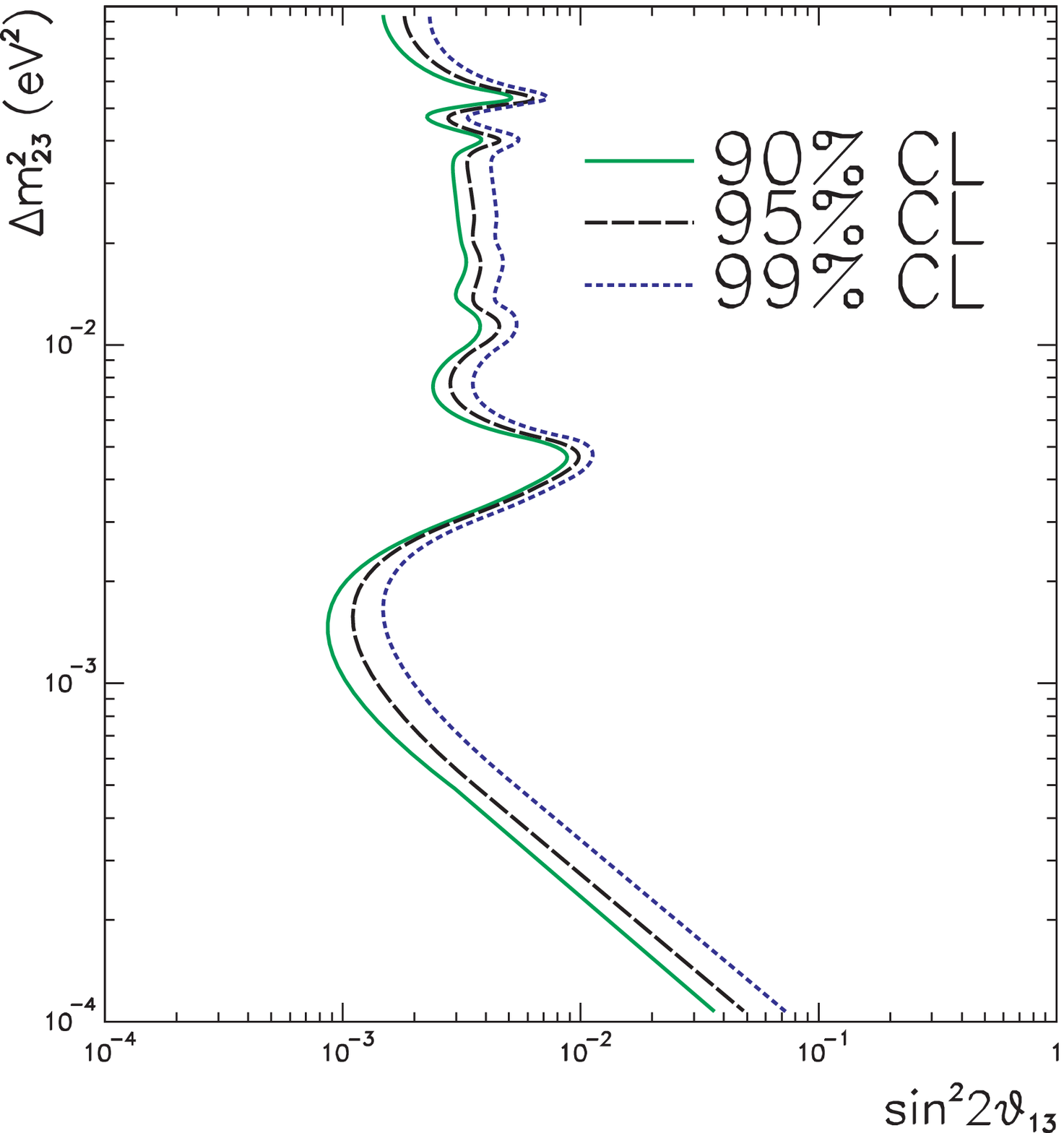}
\caption{\label{fig:sensi45}Sensitivity contours obtained with a SPL energy of $3.5$~GeV and default parameters of table~\ref{tab:param}. In particular, it is reminded that the tunnel geometry parameters are $L_T = 20$~m and $R_T = 1$~m. (\full), (\dashed) and (\dotted) curves stand for $90\%$, $95\%$ and $99\%$ confidence level, respectively.}
\end{figure}

\begin{figure}
\centering
\includegraphics[width=85mm]{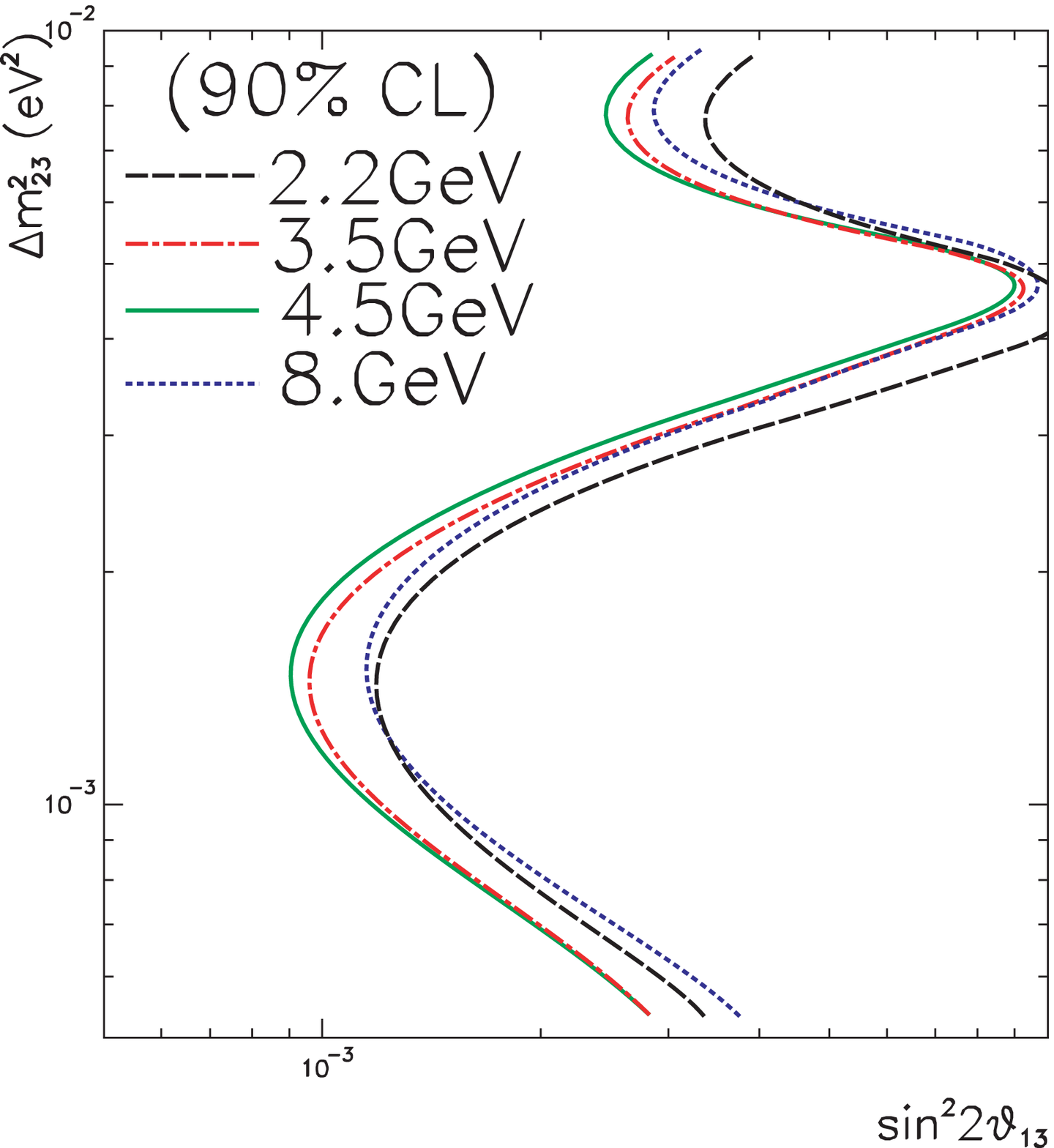}
\caption{\label{fig:compSensi}Comparison of 90\% CL sensitivity contours obtained with SPL energies of $2.2$~GeV (\dashed), $3.5$~GeV (\chain), $4.5$~GeV (\full) and $8$~GeV (\dotted) and default parameters of table~\ref{tab:param}. In particular, it is reminded that the tunnel geometry parameters are $L_T = 20$~m and $R_T = 1$~m.}
\end{figure}

The contours at $90\%$, $95\%$ and $99\%$~CL of the $\theta_{13}$ sensitivity are presented in the ($\sin^22\theta_{13}$, $\Delta m^2_{23}$) plane on figure~\ref{fig:sensi45} for $3.5$~GeV proton beam kinetic energy. 
The comparison between the contours at $90\%$~CL with $2.2$~GeV, $3.5$~GeV, $4.5$~GeV and $8$~GeV beam energies is shown on figure~\ref{fig:compSensi}. One notices in this scenario a better performance reached with a $4.5$~GeV energy beam as a confirmation of significance parameter value. But, in fact there is not much visual difference between a sensitivity obtained with $3.5$~GeV and $4.5$~GeV, even if one should keep in mind that kaon production models are different at these two energies (see section~\ref{sec:kaon}). These two energy settings have been studied with different decay tunnel geometry and  results are reported on table~\ref{tab:thvsE_3545}. One notices that similar results can be reached with a $3.5$~GeV beam, compared to a $4.5$~GeV beam.

Quantitative studies of the minimum $\sin^22\theta_{13}$ with respect to the kinetic beam energy $E_k(proton)$, and the decay length $L_T$, and the systematics $\epsilon_{syst}$ are presented in tables~\ref{tab:thvsE} and \ref{tab:thvseps}. One notices that for $\epsilon_{syst} = 5\%$ there is no difference between a $3.5$~GeV and a $4.5$~GeV beam.

\begin{figure}
\centering
\includegraphics[width=85mm]{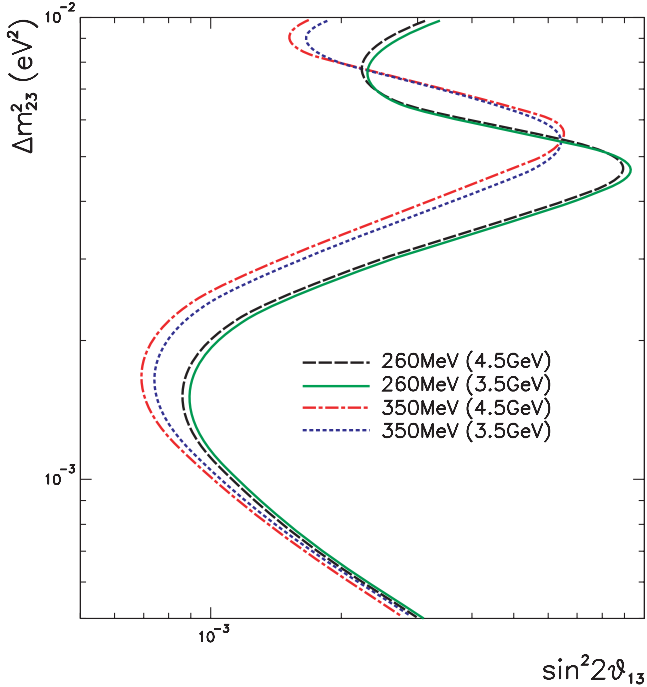}
\caption{\label{fig:comp5year}Comparison of 90\% CL sensitivity contours obtained with SPL energies of $3.5$~GeV or $4.5$~GeV, and either a $260$~MeV (default) neutrino beam or a $350$~MeV neutrino beam. The tunnel geometry parameters are $L_T = 40$~m and $R_T = 2$~m. The (\chain) curve corresponds to a $350$~MeV/$4.5$~GeV (neutrino beam/SPL beam energy) setting; the (\dotted) curve corresponds to a $350$~MeV/$3.5$~GeV setting; the (\dashed) curve corresponds to a $260$~MeV/$4.5$~GeV setting and the (\full) curve corresponds to a  $260$~MeV/$3.5$~GeV setting.}
\end{figure}

We have also considered the $3.5$~GeV and $4.5$~GeV beam energies with the tunnel geometry parameters $L_T=40$~m and $R_T=2$~m, and the horn design producing a $350$~MeV neutrino beam (see section~\ref{sec:horn}). In table~\ref{tab:thvsE_3545} are reported numerical values, and on figure~\ref{fig:comp5year} are shown the $90$\% CL sensitivity contours. With the $350$~MeV neutrino beam, one can expect a $16$\% improvment with respect to the $260$~MeV neutrino beam for the same decay geometry. One also notices that there is marginal gain to increase the beam energy from $3.5$~GeV to $4.5$~GeV, as already mentioned. 

As well, there are  variations on the minimum $\sin^22\theta_{13}$ value that may be reached in a $\nu_\mu \rightarrow \nu_e$ experiment which are due to the $\mathrm{sign}(\Delta m^2_{23})$ ambiguity and the $\delta_{CP}$ value. On table~\ref{tab:sign} are presented these kinds of variations. Other ambiguities coming from the sign$(\tan(2\theta_{23}))$ ignorance also occur as studied in reference \cite{DONINI}. From figure~9 of this reference, we estimate a 30\% effect on $\sin^2(2\theta_{13})$ sensivity due to these ambiguities.

\begin{table*}
\centering
\caption{\label{tab:thvsE_3545}Minimum $\sin^22\theta_{13}\times 10^3$ in the $(\sin^22\theta_{13},\Delta m^2_{23})$ plane observable at $90\%$ CL computed for different decay tunnel length ($L_T$) and radius ($R_T$) for the $3.5$~GeV and $4.5$~GeV scenarios and 5 years of positive focusing. Other parameters are fixed to default values (table~\ref{tab:param}). Settings in parenthesis are identical to those of table~\ref{tab:speciesfluxes}, except that the setting (7b) corresponds to the tunnel geometry of setting (7) but the horn geometry producing a $350$~MeV neutrino beam is used. We remind that the setting (2) is the default one, and the settings (7) and (7b) correspond to $L_T = 40$~m and $R_T = 2$~m.}
\begin{tabular}{@{}l*{15}{l}}
\hline\noalign{\smallskip}
setting & (1) & (2) & (3) & (4) & (5) & (6) & (7) & (7b) & (8) & (9)\\
\noalign{\smallskip}\hline\noalign{\smallskip}
$3.5$~GeV   & $0.92$ & $0.92$ & $0.83$ & $0.98$ & $1.00$ & $0.93$ & $0.91$ & $0.76$ & $1.05$ & $1.01$ \\
$4.5$~GeV   & $1.04$ & $0.89$ & $0.82$ & $0.94$ & $0.99$ & $0.92$ & $0.87$ & $0.71$ & $1.03$ & $1.00$ \\
\noalign{\smallskip}\hline
\end{tabular}
\end{table*}

\subsection{Mixed positive/negative focusing scenario}
\begin{figure}
\centering
\includegraphics[width=85mm]{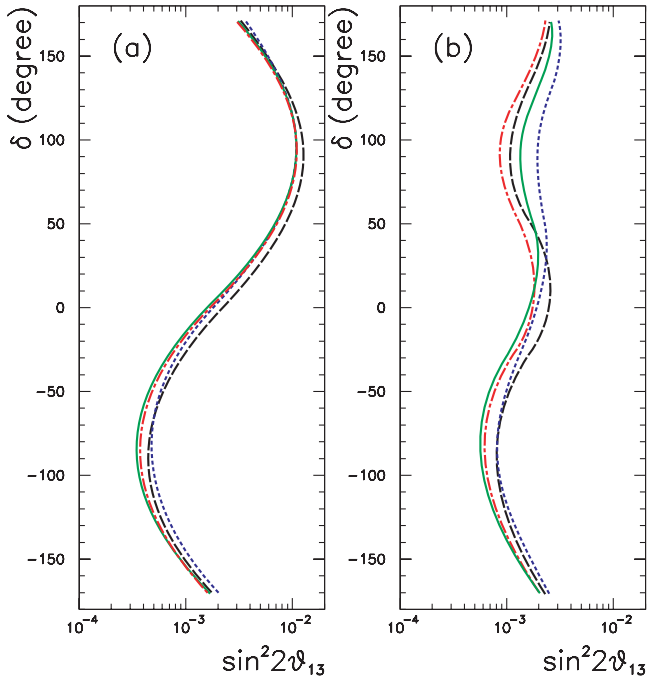}
\caption{\label{fig:compDeltaTheta}90\% sensitivity contours obtained with SPL beam energy of $2.2$~GeV (\dashed), $3.5$~GeV (\chain), $4.5$~GeV (\full) and $8$~GeV (\dotted) at $90\%$ CL. Default parameters of table~\ref{tab:param} are used either with a 5 years positive focusing scenario (a) or a mixed scenario of 2 years positive focusing and 8 years of negative focusing (b). In particular, it is reminded that the tunnel geometry parameters are $L_T = 20$~m and $R_T = 1$~m.}
\end{figure}

\begin{figure}
\centering
\includegraphics[width=85mm]{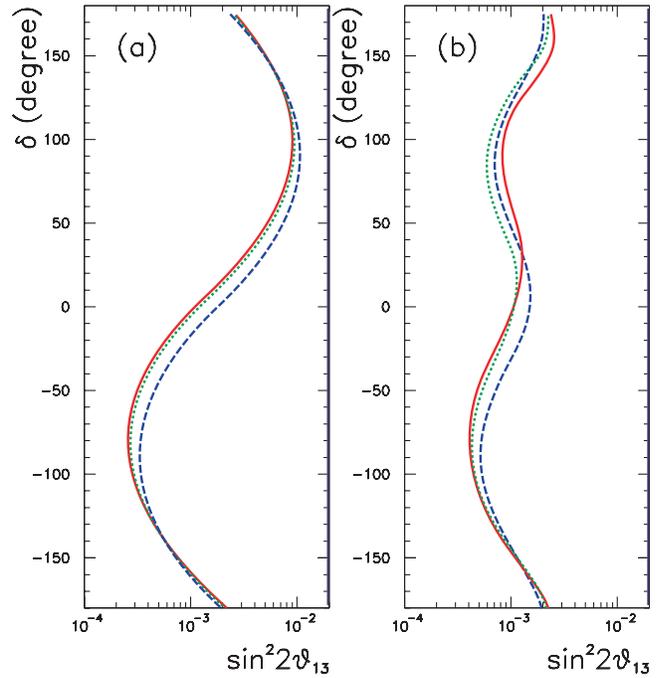}
\caption{\label{fig:sensiDeltaTheta}90\% CL sensitivity contours obtained with the decay tunnel geometry parameters $L_T=40$~m and $R_T=2$~m and different SPL beam energies ($3.5$~GeV or $4.5$~GeV) and different horn designs ($260$~MeV or $350$~MeV neutrino beams): (\full) curve for a $350$~MeV/$4.5$~GeV setting, (\dotted) curve for a $350$~MeV/$3.5$~GeV setting, (\dashed) curve for a $260$~MeV/$3.5$~GeV setting. Other default parameters of table~\ref{tab:param} are used either with a 5 years positive focusing scenario (a) or a mixed scenario of 2 years positive focusing and 8 years of negative focusing (b).}
\end{figure}

The combined $\sin^22\theta_{13}$ and $\delta_{CP}$ sensitivity for the 5 years positive focusing scenario and the default parameters of table~\ref{tab:param} is presented on figure~\ref{fig:compDeltaTheta}(a). The results obtained with a $3.5$~GeV and $4.5$~GeV SPL beam are similar and better than with the other energy settings. On figure~\ref{fig:sensiDeltaTheta}(a) the results obtained with a $260$~MeV neutrino beam and a $350$~MeV neutrino beam are presented with a $40$~m long, $2$~m radius decay tunnel. With the $350$~MeV neutrino beam, one can reached better sensitivity results in the range $|\delta_{CP}| < 120^o$, and comparatively the gain obtained when switching from a $3.5$~GeV proton beam to a $4.5$~GeV proton beam is marginal. 
\begin{table}
\centering
\caption{\label{tab:thvseps}Minimum $\sin^22\theta_{13}\times 10^3$ in the $(\sin^22\theta_{13},\Delta m^2_{23})$ plane observable at $90\%$ CL computed for different level of systematics ($\epsilon_{syst}$) and kinetic beam energy ($E_k(proton)$) and 5 years of positive focusing. Other parameters are fixed to default values (table~\ref{tab:param}).}
\begin{tabular}{@{}l*{15}{l}}
\hline\noalign{\smallskip}
			      & $2.2$~GeV & $3.5$~GeV & $4.5$~GeV & $6.5$~GeV & $8$~GeV \\
\noalign{\smallskip}\hline\noalign{\smallskip}
			$2\%$ & $1.16$ & $0.92$ & $0.89$ & $1.01$ & $1.12$ \\
			$5\%$ & $1.48$ & $1.25$ & $1.25$ & $1.48$ & $1.64$ \\
			$10\%$ & $2.40$ & $2.14$ & $2.21$ & $2.72$ & $3.09$ \\
\noalign{\smallskip}\hline
\end{tabular}
\end{table}
\begin{table}
\centering
\caption{\label{tab:sign}Minimum $\sin^22\theta_{13}\times 10^3$ in the $(\sin^22\theta_{13},\Delta m^2_{23})$ plane observable at $90\%$ CL computed for a $2.2$~GeV kinetic energy proton beam, and for different values of sign$(\Delta m^2_{23})$ and $\delta_{CP}$ and 5 years of positive focusing. Other parameters are fixed to default values (table~\ref{tab:param}).}
\begin{tabular}{@{}l*{15}{l}}
\hline\noalign{\smallskip}
			    & $-180^\circ$ & $-90^\circ$ & $0^\circ$ & $90^\circ$ & $180^\circ$\\
\noalign{\smallskip}\hline\noalign{\smallskip}
			$+$ & $1.40$ & $0.43$ & $1.16$ & $11.48$ & $1.40$\\
			$-$ & $1.45$ & $11.75$& $1.11$ & $0.43$  & $1.45$ \\
\noalign{\smallskip}\hline
\end{tabular}
\end{table}
To improve the $\delta_{CP}$-independent limit on $\sin^22\theta_{13}$, especially around $\delta_{CP}=90^\circ$, one may envisage a combination of 2 years with positive focusing and 8 years negative focusing as in references \cite{DONINI,JJG,Mezzetto}. The comparison of the results obtained with different SPL beam energies on the combined sensitivity contours are presented in figure~\ref{fig:compDeltaTheta}(b). Quantitative results with this kind of mixed focusing scenario are reported table \ref{tab:thvsE_td}. One generally gets $10\%$ to $20\%$ better limit on $\sin^22\theta_{13}$ independently of $\delta_{CP}$ with a $3.5$~GeV kinetic energy beam compared to a $2.2$~GeV beam. Doubling the length and the radius of the decay tunnel allows to reach a $10\%$ better limit.

On figure~\ref{fig:sensiDeltaTheta}(b) are presented the results considering the effects of a $350$~MeV neutrino beam obtained either with a $3.5$~GeV proton beam or a $4.5$~GeV proton beam compared to a $260$~MeV neutrino beam obtained with a $4.5$~GeV proton beam. The tunnel geometry parameters are $L_T=40$~m and $R_T=2$~m (other tunnel geometry have been studied but the results are worse and so are not reported). Except in the region $|\delta_{CP}|>150^o$, the results obtained with the $350$~MeV neutrino beam ($3.5$~GeV proton beam) are somewhat better, even if a $11\%$ improvement of the $\delta_{CP}$-independent $\sin^22\theta_{13}$ limit can be reached with the $260$~MeV neutrino beam obtained with the $3.5$~GeV proton beam.
\begin{figure}
\centering
\includegraphics[width=85mm]{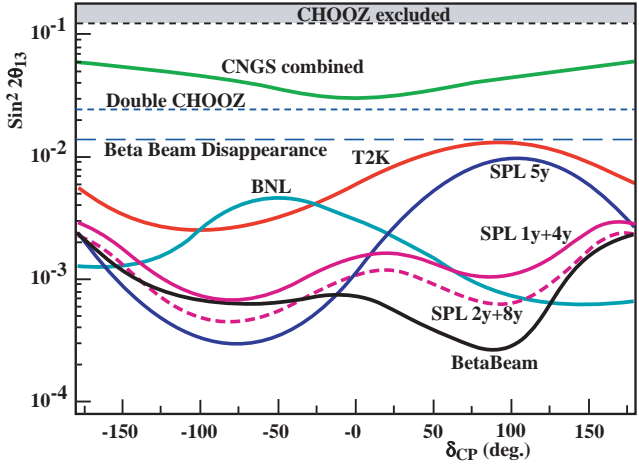}
\caption{\label{fig:deltathetaFinal}90\%CL sensitivity contours labeled by the project or experiment involved. The "CHOOZ excluded" dashed curve comes from the exclusion obtained from reference \cite{CHOOZ} with $\Delta m^2 = \Delta m^2_{atm}$; in the same conditions is given the sensitivity foreseen for the "Double-CHOOZ" project \cite{Wpaper}. The "CNGS combined" has been obtained combining the results form OPERA and ICARUS \cite{MIGLIOZZITERANOVA}. The T2K contour has been derived from reference \cite{KOBAYASHI}. The BNL contour has been obtained from reference \cite{BNL}. The "Beta Beam" contour has been computed with 5 years running with both $\nu_e$ and $\bar{\nu}_e$ neutrino beams in an appearance mode, while the dashed "Beta Beam disappearance" has been obtained as if the $\beta$ beam were analysed like a reactor experiment with 1\% systematic error \cite{MAOROPRIVATE}. The "SPL 5y" and "SPL 2y+8y" and "SPL 1y+4y" curves have been obtained from the optimisation described in this paper ("5y": positive only focusing scenario; "1y+4y": 1 year of positive focusing and 4 years of negative focusing scenario; "2y+8y": 2 years of positive focusing and 8 years of negative focusing scenario) using a $3.5$~GeV beam and a decay tunnel of $40$~m length, and $2$~m radius.}
\end{figure}
\section{Summary and outlook}
A complete chain of simulation has been set up for the SPL-Fréjus project. The neutrino production has been extended to the kaon decay contribution, which is important to test SPL energy scenario above $2.2$~GeV.

The beam line optimization has been performed using the sensitivity to $\sin^22\theta_{13}$ and $\delta_{CP}$. The shape of the focusing system has been updated to obtain a neutrino beam energy around $260$~MeV or $350$~MeV.
\begin{table*}
\centering
\caption{\label{tab:thvsE_td}Minimum $\sin^22\theta_{13}\times 10^3$ observable at $90\%$ CL computed for the worse $\delta_{CP}$ case, and for different decay tunnel length ($L_T$) and radius ($R_T$) for the $3.5$~GeV and $4.5$~GeV scenarios and 2 years of positive focusing plus 8 years of negative focusing. Other parameters are fixed to default values (table~\ref{tab:param}). Settings are identical to those of table~\ref{tab:speciesfluxes}.}
\begin{tabular}{@{}l*{15}{l}}
\hline\noalign{\smallskip}
setting & (1) & (2) & (3) & (4) & (5) & (6) & (7) & (7b) & (8) & (9)\\
\noalign{\smallskip}\hline\noalign{\smallskip}
$2.2$~GeV &        & $2.52$ &        &        & $2.58$ &        & $2.30$ &        &         &       \\
$3.5$~GeV & $2.34$ & $2.22$ & $2.10$ & $2.13$ & $2.09$ & $2.08$ & $2.02$ & $2.28$ &  $2.16$ & $2.09$\\
$4.5$~GeV & $2.91$ & $2.60$ & $2.43$ & $2.48$ & $2.52$ & $2.39$ & $2.34$ & $2.55$ &  $2.53$ & $2.47$\\
\noalign{\smallskip}\hline
\end{tabular}
\end{table*}

In a positive only focusing scenario, the best limit on $\sin^22\theta_{13}$ is $0.71\times10^{-3}$
 ($90\%$ CL, $\delta_{CP}=0$), with a $4.5$~GeV beam energy and a $40$~m long, $2$~m radius decay tunnel, and a beam energy around 350 MeV. However, the $3.5$~GeV beam may also obtain rather similar limit with $0.76\times10^{-3}$ ($90\%$ CL) with the same tunnel parameters. The $\delta_{CP}$ independant $\sin^22\theta_{13}$ sensitivity is limited to $\approx 10^{-2}$ due to the $\delta_{CP} \approx 90^o$ region.
 
In a mixed focusing scenario, the best limit on $\sin^22\theta_{13}$ independent of $\delta_{CP}$ is $2.02\times10^{-3}$ ($90\%$ CL) obtained with a $3.5$~GeV beam energy, a $40$~m long, $2$~m radius decay tunnel and a beam energy around 260 MeV. But for $\vert\delta_{CP}\vert < 150^{\circ}$, the 350 MeV neutrino beam is better, keeping the primary proton energy at 3.5 GeV and with the same decay tunnel parameter.

The comparison of the optimization presented in this paper with the results obtained by other projects is displayed on figure \ref{fig:deltathetaFinal}. It presents the 5 years positive focusing scenario, and two versions of a mixed scenario using positive and negative focusing: one scenario duration is 5 years in total and the other one is 10 years running in total and has been used in the previous section. It shows the complementarity of the SPL-Fr\'ejus project with the beta beam-Fr\'ejus project. Especially when considering the sensitivity to $\sin^22\theta_{13}$ for $\delta_{CP}<0$. 

The authors think that the present study may be extended in many respects. The beam line simulation part may be performed with a single simulator as FLUKA (or GEANT4 \cite{GEANT4} for comparison). Other targets may be envisaged (tantalum, carbon) as well as other detector types as a Large Liquid Argon detector \cite{BIGICARUS}. The baseline length may also be revisited as well as the off axis option. The sensitivity analysis may be deeper investigated using the complete set of possible ambiguities as in reference \cite{DONINI}, and the $\theta_{13}$ or $\delta_{CP}$ measurement accuracy with new beam energy scenario may be investigated too.     
\begin{acknowledgement}
The authors would like to thank M.~Mezzetto for expressing his interest since the early stage of this work and for providing us with his sensitivity computation program. Also the authors thank S.~Gilardoni for fruitful discussions.
\end{acknowledgement}
\appendix
\setcounter{section}{0}
\section{Decay probability computations}
\label{sec:decayprobcomp}
This appendix contains the probability formulas and the algorithms used in the flux computation (see section~\ref{sec:algo}).

\subsection{Pion neutrino probability computation}
\label{sec:Ppi}
\begin{figure}
\centering
\includegraphics{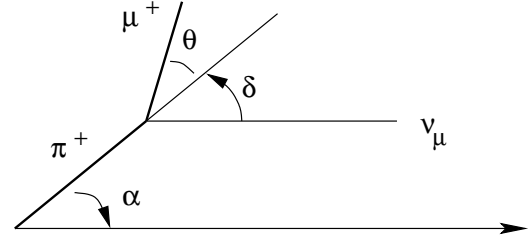}
\caption{\label{fig:pionDecay}Pion decay in the tunnel frame. To reach the detector, $\delta = -\alpha$ is needed.}
\end{figure}

Pions decay only as $\pi^+\rightarrow \mu^+ + \nu_\mu$ or $\pi^- \rightarrow \mu^- + \bar{\nu}_\mu$ and the neutrinos are emitted isotropically in the pion rest frame, with an energy of about $30$~MeV given by the 2-body decay kinematics. Applying a Lorentz boost knowing the pion momentum and direction, it is possible to compute the probability to reach for the neutrinos the detector. Only neutrinos parallel to the beam axis are supposed to pass through the detector fiducial area, and therefore, the neutrinos must be emitted by the pion with an angle opposite to the angle between the pion and the beam axis (see figure~\ref{fig:pionDecay}). This gives:
\begin{equation}
\mathcal{P}_\pi = \frac{1}{4\pi}\frac{A}{L^2}\frac{1-\beta^2}{(\beta\cos\alpha-1)^2}
\label{probaPi}
\end{equation}
where $\beta$ is the velocity of the pion in the tunnel frame, $A$ is the fiducial detector surface, $L$ the distance between the neutrino source and the detector, and $\alpha$ the angle between the pion direction and the beam axis in the laboratory frame.

\subsection{Muon neutrino probability computation}
\label{sec:Pmu}
Muons decay only as $\mu^+ \rightarrow e^+ + \nu_e + \bar{\nu}_\mu$ or $\mu^- \rightarrow e^- + \bar{\nu}_e + \nu_\mu$, and will produce background events. The mean decay length of the muons is $2$~km, therefore, most of them do not decay in the tunnel. This induces a lack of statistics to estimate the corresponding level of background. This problem has been solved using each muon appearing in the simulation in the following steps:
\begin{enumerate}
	\item the probability for the muon to decay into the tunnel has been computed using a straight line propagation;
	\item the available energy for the neutrino in the tunnel frame has been divided in $20$~MeV energy bins; 
	\item one $\nu_e$ and one $\nu_\mu$ have been simulated in each of the energy bins (step 2). Then, the probability to reach the detector has been computed, and multiplied by the probability computed at step (1).
\end{enumerate}
After the probability computation, the non useful muon is discarded by GEANT to gain in CPU time. 

The probability for the muon neutrino and the electron neutrino to be emitted parallel to the beam axis is \cite{donega}:
\begin{multline}
\frac{d\mathcal{P}_\mu}{dE_\nu} = \frac{1}{4\pi}\frac{A}{L^2}\frac{2}{m_\mu}\frac{1}{\gamma_\mu(1+\beta_\mu\cos\theta^*)}\\ \times \frac{1-\beta_\mu^2}{(\beta_\mu\cos\rho-1)^2}\left[f_0(x)\mp \Pi_\mu^L f_1(x)\cos\theta^*\right]
\label{probaMu}
\end{multline}
where $\beta_\mu$ and $\gamma_\mu$ are the velocity and the Lorentz boost of the muon in the tunnel frame, $\theta^*$ is the angle with respect to the beam axis of the muon in the muon rest frame, $\rho$ is the corresponding angle in the tunnel frame. As the pion case, this angle appears because the neutrino must be parallel to the beam axis. $\Pi_\mu^L$ is the muon longitudinal polarization, the parameter $x$ is defined as $x=2E_\nu^*/m_\mu$ where $E_\nu^*$ is the neutrino energy in the muon rest frame, and the functions $f_0(x)$ and $f_1(x)$ coming from the matrix element of the muon decays are given in table~\ref{tab:Function}. The sign in front of $\Pi_\mu^L$ in equation~\ref{probaMu} is $(-)$ for the $\mu^+$ decays and $(+)$ for the $\mu^-$ decays, respectively.

\begin{table}
\centering
\caption{\label{tab:Function}Flux function in the muon rest frame \cite{Gaisser}.}
\begin{tabular}{@{}l*{15}{l}}
\hline\noalign{\smallskip}
		      &  $f_0(x)$ & $f_1(x)$ \\
\noalign{\smallskip}\hline\noalign{\smallskip}
		$\nu_\mu$ & $2x^2(3-2x)$ & $2x^2(1-2x)$ \\
		$\nu_e$   & $12x^2(1-x)$ & $12x^2(1-x)$ \\
\noalign{\smallskip}\hline
\end{tabular}
\end{table}

Muon polarization is computed using the conservation of the transverse component of the velocity four-vector $\gamma(1,\beta)$ between the muon rest frame (where the polarization is computed) and the pion rest frame, where the muon helicity is $-1$, due to the parity non conservation. It yields \cite{picasso}:
\begin{equation}
\Pi_\mu^T = \frac{\gamma_\pi\beta_\pi}{\gamma_\mu\beta_\mu}\sin\theta^*
\mbox{ and }\Pi_\mu^L = \sqrt{1-\Pi_\mu^{T2}}
\label{pola}
\end{equation}
where $\gamma_\pi$, $\beta_\pi$, $\gamma_\mu$, and $\beta_\mu$ are the Lorentz boost and velocity of the pion and of the muon in the tunnel frame, and $\theta^*$ the angle with respect to the beam axis of the muon in the pion rest frame.

\subsection{The treatment of the kaons}
\label{sec:kaons}
Contrary to pions and muons, kaons have many decay channels. They are summarized in table~\ref{tab:BRKP0SL}.

\begin{table}
\centering
\caption{\label{tab:BRKP0SL}Charged and neutral kaon decay channels \cite{pdg}.}
\begin{tabular}{@{}l*{15}{l}}
\hline\noalign{\smallskip}
\centre{2}{$K^\pm$}         & \centre{2}{$K^0_L$}     & \centre{2}{$K^0_S$}\\
\noalign{\smallskip}\hline\noalign{\smallskip}
$\mu^\pm\nu_\mu$ & $63.51\%$ & $\pi^-e^+\nu_e$ & $19.35\%$        & $\pi^+\pi^-$ & $68.61\%$ \\
$\pi^\pm\pi^0$ & $21.17\%$   & $\pi^+e^-\bar{\nu}_e$ & $19.35\%$  & $\pi^0\pi^0$ &  $31.39\%$ \\ 		
$\pi^\pm\pi^+\pi^-$ & $5.59\%$ & $\pi^-\mu^+\nu_\mu$ & $13.5\%$ & & \\
$e^\pm\nu_e\pi^0$ & $4.82\%$   & $\pi^+\mu^-\bar{\nu}_\mu$ & $13.5\%$ & & \\  			  			
$\mu^\pm\nu_\mu\pi^0$ & $3.18\%$ & $\pi^0\pi^0\pi^0$ & $21.5\%$ & &\\
$\pi^\pm\pi^0\pi^0$ & $1.73\%$ & $\pi^+\pi^-\pi^0$ & $12.38\%$ & & \\ 			
\noalign{\smallskip}\hline
\end{tabular}
\end{table}

There is a very small amount of kaons produced (section~\ref{sec:kaon}), and this number has been artificially increased in order to obtain statistically significant results.  The multiplicity of decay channels makes impossible the method used for the muon case (\ref{sec:Pmu}). The method chosen for the good compromise between the gain in CPU and the statistical uncertainty of the results, is to duplicate many times each kaon exiting the target. The number of duplication varies between 10 and 300. It depends on the initial kaon rate and therefore on the beam energy.

All the kaons daughter particles are tracked by GEANT until they decay. Three different types of daughter particles are identified in the kaon decays. The first type corresponds to primary neutrinos, the second type concerns charged pions and muons, and the neutral pions are left for the last type.

In the $K^\pm\rightarrow\mu^\pm \nu_\mu(\bar{\nu}_\mu)$ decay modes, the computation of the probability for a neutrino to reach the detector is the same than the 2-body decay formula used to in the pion decay (equation~\ref{probaPi}), where $\beta$ is now the kaon velocity, and $\alpha$ the angle of the kaon with respect to the beam axis. 

When a neutrino is produced by a kaon 3-body decay, the probability to reach the detector is computed using a pure phase space formula. It yields:
\begin{multline}
\frac{d\mathcal{P}_K}{dE_\nu} = \frac{1}{4\pi}\frac{A}{L^2}\frac{1}{m_K-m_\pi-m_l} \\ \times \frac{1}{\gamma_K(1+\beta_K\cos\theta^*)}
\frac{1-\beta_K^2}{(\beta_K\cos\delta-1)^2}
\label{probaL}
\end{multline}
where $m_K$ is the kaon mass (charged or neutral), $m_\pi$ is the pion mass ($\pi^0$ mass in $K^\pm$ decays and $\pi^\pm$ mass in $K_L^0$ decays), and $m_l$ is the mass of the lepton associated with the neutrino. The $\beta_K$ and $\gamma_K$ are the velocity and the Lorentz boost of the kaon, $\theta^*$ is the angle between the neutrino direction and the kaon direction, in the kaon rest frame. Finally, $\delta$ is the angle between the kaon direction and the beam axis in the tunnel frame.

When a $\pi^\pm$ is produced in the kaon decay chain, it is tracked by GEANT until it decays, and the probability of equation~\ref{probaPi} is applied to the produced neutrino. In case of a muon, it is treated as explained in \ref{sec:Pmu}. The muon polarization is computed this time using the kaon decay informations. Finally, when a $\pi^0$ is produced, as it cannot create neutrinos, it is simply discarded.

%

%
\end{document}